\documentclass[a4paper,11pt]{article}
\pdfoutput=1 

\usepackage{jcappub} 

\usepackage[T1]{fontenc} 

\usepackage{amsmath}
\usepackage{xcolor}
\usepackage{graphicx}
\usepackage{mathtools}
\usepackage[normalem]{ulem}
\usepackage{appendix}
\usepackage{verbatim}
\usepackage{array}

\usepackage{custom_commands}

\newcommand{\vecI}{\mbox{\boldmath $I$}}
\newcommand{\vecQ}{\mbox{\boldmath $Q$}}
\newcommand{\vecU}{\mbox{\boldmath $U$}}
\newcommand{\vecn}{\mbox{\boldmath $n$}}
\newcommand{\vecs}{\mbox{\boldmath $s$}}
\newcommand{\vecm}{\mbox{\boldmath $m$}}
\newcommand{\vecd}{\mbox{\boldmath $d$}}
\newcommand{\veclambda}{\mbox{\boldmath $\lambda$}}
\newcommand{\vecP}{\mbox{\boldmath $P$}}
\newcommand{\vecu}{\mbox{\boldmath $u$}}

\newcommand\tomo[1]{{\color{purple} #1}}

\def\LB{\textit{LiteBIRD }}
\def\LBm{\textit{LiteBIRD}}
\def\Planck{\textit{Planck}}

\def\gtorder{\mathrel{\raise.3ex\hbox{$>$}\mkern-14mu
             \lower0.6ex\hbox{$\sim$}}}
\def\ltorder{\mathrel{\raise.3ex\hbox{$<$}\mkern-14mu
             \lower0.6ex\hbox{$\sim$}}}

\definecolor{mygreen}{RGB}{0, 100, 0}

\title{Effect of Instrumental Polarization with a Half-Wave Plate on the $B$-Mode Signal: Prediction and Correction}

\author[a]{Guillaume Patanchon}
\author[b]{Hiroaki Imada}
\author[c]{Hirokazu Ishino}
\author[d,e]{Tomotake Matsumura}

\affiliation[a]{Université Paris-Cité, CNRS, Astroparticule et Cosmologie, F-75013 Paris, France}
\affiliation[b]{National Astronomical Observatory of Japan, Mitaka, Tokyo 181-8588, Japan}
\affiliation[c]{Okayama University, Department of Physics, Okayama 700-8530, Japan}
\affiliation[d]{Kavli Institute for the Physics and Mathematics of the Universe (Kavli IPMU, WPI), UTIAS, The University of Tokyo, Kashiwa, Chiba 277-8583, Japan}
\affiliation[e]{Center for Data-Driven Discovery, Kavli IPMU (WPI), UTIAS, The University of Tokyo, Kashiwa, Chiba 277-8583, Japan}

\abstract{
We evaluate the effect of half-wave plate (HWP) imperfections inducing intensity leakage to the measurement of Cosmic Microwave Background (CMB) $B$-mode polarization signal with future satellite missions focusing on the tensor-to-scalar ratio $r$. The HWP is modeled with the Mueller formalism, and coefficients are decomposed for any incident angle into harmonics of the HWP rotation frequency due to azimuthal angle dependence. Although we use a general formalism, band-averaged matrix coefficients are calculated as an example for a 9-layer sapphire HWP using EM propagation simulations. We perform simulations of multi-detector observations in a band centered at 140\,GHz using \LB instrumental configuration. We show both theoretically and with the simulations that most of the artefacts on Stokes parameter maps are produced by the dipole leakage on $B$-modes induced by the fourth harmonics $M^{(4f)}_{QI}$ and $M^{(4f)}_{UI}$. The resulting effect is strongly linked to the spin-2 focal plane scanning cross linking parameters. We develop a maximum likelihood-based method to correct the IP leakage by joint fitting of the Mueller matrix coefficients as well as the Stokes parameter maps. 
We show that the residual leakage after correction leads to an additional noise limited uncertainty on $r$ of the order of $10^{-7}$, independently of the value of the Mueller matrix coefficients. We discuss the impact of the monopole signal and the potential coupling with other systematic effects such as gain variations and detector nonlinearities. 

}


\begin{document}

\maketitle
\flushbottom

\section{Introduction}

Cosmic Microwave Background (CMB) radiation polarization is a powerful probe of fundamental physics and astrophysics in the early Universe. The measurement of the CMB temperature fluctuations and polarization has allowed accurate determination of 6 parameters of the standard cosmological model~\cite{PlanckCosmo18} and reinforced the hypothesis of an inflationary period of the early Universe~\cite{PlanckInflation18} at the origin of perturbations. The tensor modes generated during the inflation, which remain undetected today, imprint a distinct negative parity pattern of the CMB polarization distribution on the sky called $B$-modes~\cite{Seljak97}. The measurement of the power spectrum of the $B$-modes will give access to the physics of the inflation and particularly the determination of its energy scale in the most simple scenarios~\cite{Liddle94}. Many experiments are targeting the $B$-mode signal either from the ground or from space with the planned satellite mission \LB of JAXA~\cite{LB2021}. This measurement is challenging because of the presence of foreground components, mainly polarized Galactic emission at sub-millimetre wavelengths, and of E-mode transfer to $B$-modes due to gravitational lensing of CMB photons by large scale structures in the Universe. Astrophysical components dominate over the CMB polarization signal~\cite{BicepKeckPlanck15}, especially at large angular scales, at the multipoles of the reionisation bump in the $B$-mode power spectrum, while lensing dominates at smaller scales, over the recombination bump for a tensor-to-scalar ratio $r \lesssim 10^{-2}$~\cite{PolarbearLensB}. Current best constraints are obtained with the BICEP/Keck experiment in combination with \Planck\ and \textit{WMAP} data giving $r<0.036$ at 95\,\% confidence~\cite{BicepKEK21}. Future projects in the next decade are targeting the sensitivity of $r < 10^{-3}$.
Component separation and modelling of Galactic foregrounds is a crucial step for reaching this goal.

Moreover, polarization measurement is inherently subject to several systematic effects because it requires differencing the signal from more than two measurements (see e.g.~\cite{Couchot99}), either by using data from the same detector at very different time intervals and/or from different detectors (usually a pair of detectors with orthogonal polarizer orientations is used). Small mismatches between measurements, which can be due to beam mismatches, gain or band-pass mismatches are introducing leakage from intensity to polarization in the presence of large amplitude unpolarized background signal~\cite{Rosset07,BPMM}.
Measurement mismatches are strongly reduced by the use of a rapidly rotating HWP allowing quick modulation of the polarization signal by rotating the polarization angle at 4 times the rotation frequency of the HWP, and hence allows quasi-instantaneous measurement of Stokes parameters by each individual detector. 
However HWPs are not perfect optical devices even assuming an ideal design (without manufacture imperfections) and optimal configuration of layers and anti-reflection coating in case of a transmissive HWP (see e.g~\cite{Kunimoto21}).
Transmissive HWP are usually made broad band and are optimized for a specified frequency range by combining several layers~\cite{Pancharatnam55}.
\LB will use HWPs as first optical elements, one for each of the three instruments, the Low-Frequency Telescope (LFT), the Mid-Frequency Telescope (MFT) and the High-Frequency Telescope (HFT), and is then less subject to optics imperfections such as instrumental polarization seen in other ground and balloon experiments using HWPs, see~\cite{PolarBearHWP17,EBEXHWP19}. As a satellite, it is also functioning in a more stable environment without large fluctuations from atmospheric emission leading to a more predictable intensity to polarization signal and to less coupling of the effects of HWP imperfections with other systematic effects such as gain variations as discussed in~\cite{EBEXHWP19}. The effects of HWP imperfections for ground experiments using a HWP at the entrance aperture are also described in~\cite{Essinger16b}.

In this paper we evaluate the impact of imperfections (deviations from an ideal HWP) on the polarization power spectra, using the Mueller matrix formalism and by modeling the coefficients of the matrix as the sum of terms varying with the azimuthal angle around the HWP for oblique incidence, each of the terms being harmonics of the HWP rotation frequency. The main systematic effect is the instrumental polarization induced by the 4$f_{\rm HWP}$ terms, i.e. varying with respect to the azimuthal angle around the rotation axis of the HWP with a period of $\pi/2$. This effect is mostly caused by the dipole signal (and by the monopole) and to some extent by the intensity Galactic emission leaking into the estimated $E$- and $B$-mode polarization signals. The study is restrained to the 140-GHz band of LFT, at the edge of the functioning band of the transmissive HWP used for the instrument. Realistic Mueller matrix coefficients, used in this paper, are calculated and modelled in a companion paper~\cite{Imada19}, paper-I hereafter, by using EM wave propagation simulations for realistic anti-reflection coating of the HWP surface, for several observation frequencies and with respect to incident and azimuthal angles on the HWP.
We consider the case of a transmissive HWP but the formalism developed in this paper can be generalized to reflective HWP.
Data simulations are performed for a \LBm-type experiment for several configurations of detector locations in the focal plane. Errors due to HWP imperfections are propagated to the polarization maps using a GLS map-making code. The second part of the paper is devoted to the correction of the instrumental polarization effect at the map-making level and a method is developed. We focus on the resulting bias on the tensor-to-scale ratio after correction to evaluate the performance. Another approach using the Jones matrix formalism addressing the Q/U mixing (but not the instrumental polarization) is detailed in~\cite{Giardiello21}. As we model the full Mueller matrix in our approach, the effect of Q/U mixing is also addressed in our paper, however we give emphasis to the instrumental polarization. Regarding this aspect of Q/U mixing, a modified component separation method has been proposed in~\cite{Verges21} using a simple but extensible model of the HWP, opening the way for corrections with component separation techniques.

After introducing the observation model in Section~\ref{sec:Model}, 
the effect of intensity fluctuation leakage to the polarization including the dipole is studied in Section~\ref{sec:Prediction}. We derive an analytical prediction of the leakage effect in the maps in the same section. The effect of the combination of different detectors observation is studied in Section~\ref{sec:multi-det} and optimal configurations of detector locations are discussed.
We propose a method for correction of the effect using the dipole signal and derive to which accuracy the Mueller matrix parameters can be recovered in Section~\ref{sec:correction}.

\section{Model}
\label{sec:Model}

In this section we derive a model of the observed data with a continuously rotating HWP as a first optical element. Let us first derive the instrument model.

\subsection{Instrument and Data Model}

\begin{figure}
  \centering
  \includegraphics[width=0.59\columnwidth]{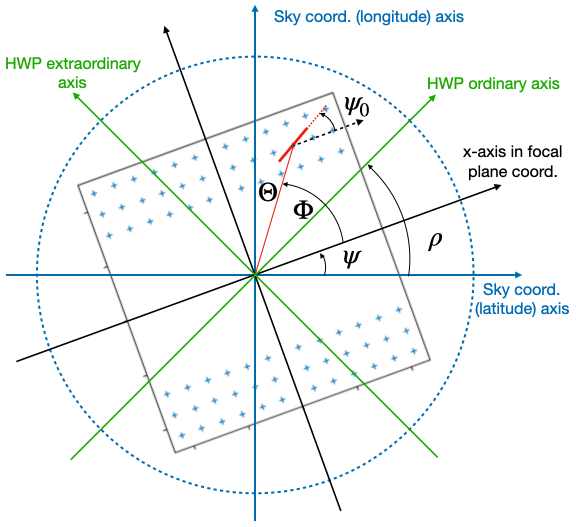}
  \includegraphics[width=0.4\columnwidth]{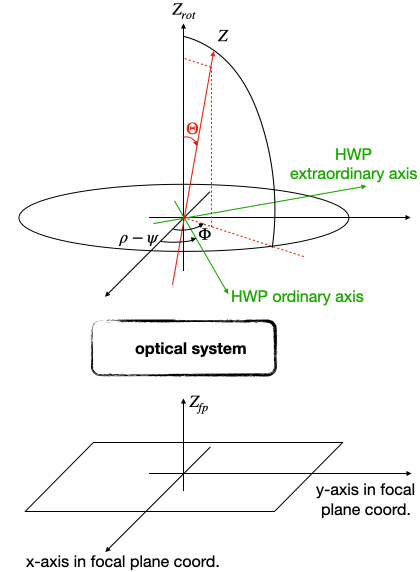}
  \caption{Instrument model and definition of the angles used throughout the paper in flat sky approximation. Crosses represent the detector pair locations. A single polarizer in the focal plane is represented to avoid confusion. The angle $\Phi$ is assumed to be zero to calculate the HWP Mueller matrix coefficients with RCWA simulations, and is used in Section~\ref{sec:multi-det} to express the coefficients at other locations in the focal plane accounting for multi-detector observations. In this illustration, the HWP optical axis is aligned with the boresight pointing while we also consider a tilted HWP in the paper.}
    \label{defInstr}
\end{figure}

We use the flat sky approximation throughout the paper.
Under this approximation we write the data observation model at a given time using the Mueller matrix formalism relating the Stokes parameters on the sky $I,Q,U,V$ to the observed signal $s$ and including all the relevant optical elements:
\begin{equation}
  s = \Big(1 \,\,\cos 2\psi_0 \,\,\sin 2\psi_0 \,\, 0 \Big)\, M(\Theta, \rho - \psi)\, R(2\psi)
\begin{pmatrix}
I \\ Q \\ U \\ V
\end{pmatrix} + n
\label{eq:modelobs}
\end{equation}
where the vector $(1 \,\,\cos 2\psi_0 \,\,\sin 2\psi_0 \,\, 0)$ models a polarizer at angle $\psi_0$ with respect to a reference axis in the focal plane coordinates; $\psi$ is the rotation angle of the reference axis of the focal plane (FP) with respect to the sky reference frame; $\rho=\omega_{\rm HWP}\,t + \rho_0$ the HWP rotation angle with respect to the sky reference frame\footnote{The rotation axis of the HWP is perpendicular to its ordinary and extraordinary axis.}; $\Theta$ is the incident angle with respect to the rotation axis of the HWP; $M(\Theta,\rho-\psi)$ is the Mueller matrix of the HWP in the focal plane coordinates, which is a 4 by 4 matrix:
\begin{equation}
M = \begin{pmatrix}
M_{II} & M_{IQ} & M_{IU} & M_{IV} \\
M_{QI} & M_{QQ} & M_{QU} & M_{QV} \\
M_{UI} & M_{UQ} & M_{UU} & M_{UV} \\
M_{VI} & M_{VQ} & M_{VU} & M_{VV}
\end{pmatrix};
\end{equation}
finally $R(2\psi)$ is the rotation matrix for Stokes parameters:
\begin{equation}
 R(2\psi) = \begin{pmatrix}
1 & 0 & 0 & 0\\
0 & \cos 2\psi & \sin 2\psi & 1\\
0 & -\sin 2\psi & \cos 2\psi & 1\\
0 & 0 & 0 & 1
\end{pmatrix}.
\end{equation}
For an ideal HWP, the Mueller matrix is:
\begin{equation}
M^{\rm Ideal} = \begin{pmatrix}
1 & 0 & 0 & 0 \\
0 & \cos 4(\rho-\psi) & \sin 4(\rho-\psi) & 0 \\
0 & \sin 4(\rho-\psi) & -\cos 4(\rho-\psi) & 0 \\
0 & 0 & 0 & -1
\end{pmatrix}.
\label{Mideal}
\end{equation}
The definition of angles used in this paper is illustrated by Figure~\ref{defInstr}.
In our model, the signal $(I,Q,U,V)^t$ is the detector beam convolved signal, and we assume a possible multiplication of $(I,Q,U,V)^t$ and $M$ terms.  This implies that there is no interference between the beams and the HWP and that the properties of the HWP do not vary significantly over the extent of the main beam. This is justified by the slow variations of the HWP Mueller matrix with the incident angles ($\Theta$, $\phi$) as we will see in the following paragraphs. The mixed effects of the HWP imperfections and the beams are studied in~\cite{Essinger16} and in~\cite{Duivenvoorden21} in the context of polarized beams. We also assume that the beam are symmetric such that their effect can be accounted at the map level before any timestream simulation (no $\psi$ angle dependence of $(I,Q,U,V)^t$). The coupling of the sidelobe systematic effect with HWP imperfections would deserve a separate study and is postponed to a future publication. In a same way, band effects are not included in our model and as a simplification we factorize the band-averaged Mueller matrix $M$ from the sky signal. In reality, because of the different emission spectra of the different components, the averaged Mueller matrix would be different for each components (this has been quantified in paper-I and is shown to be a small effect for our study). The effect would become relevant in case of a multi-frequency band study. The colour effects in case of normal incidence are studied in~\cite{Giardiello21}.

We use the Mueller matrix coefficients estimated from electro-magnetic wave propagation described in paper-I using Rigorous Coupled-Wave Analysis (RCWA)~\cite{RCWA95} simulation algorithm. According to this study
the Mueller matrix can be decomposed into a limited number of terms corresponding to different harmonics of the rotation frequency. Those terms depend on the azimuthal angle of the HWP in the frame of the focal plane $\rho-\psi$, in which detectors are fixed. The expanded Mueller matrix writes:
\begin{equation}
M_{ij}(\Theta,\rho - \psi) = M_{ij}^{(0f)}(\Theta)\,+\,M_{ij}^{(2f)}(\Theta,2\rho - 2\psi)\,+\,M_{ij}^{(4f)}(\Theta,4\rho - 4\psi)+...,
\end{equation}
with $M^{(0f)}$ a constant matrix with respect to the azimuthal angle; a term varying at twice the rotation frequency:
\begin{equation}
 M_{ij}^{(2f)}(\Theta,2\rho-2\psi)=[M_0]^{(2f)}_{ij}(\Theta)\cos (2\rho - 2\psi + \phi_{ij}^{(2f)});
 \label{M2f}
\end{equation}
and a term varying at 4 times the rotation frequency:
\begin{equation}
M_{ij}^{(4f)}(\Theta,4\rho-4\psi)=[M_0]^{(4f)}_{ij}(\Theta)\cos (4\rho - 4\psi + \phi_{ij}^{(4f)}).
\label{M4f}
\end{equation}
The Mueller matrix elements $[M_0]^{(0f)}_{ij}$, $[M_0]^{(2f)}_{ij}$, $[M_0]^{(4f)}_{ij}$, as well as the phases $\phi_{ij}^{(2f)}$ and $\phi_{ij}^{(4f)}$ are calculated in paper-I for the studied band. Because the $M^{(4f)}$ matrix varies at the same frequency than the modulated $Q$ and $U$ parameters in data, this matrix has the largest impact on the instrumental polarization induced bias. Higher order terms are found to be small and their impact on the sky maps is limited since they do not project efficiently into the map Stokes parameters.
For the following let's define $\epsilon_1$ and $\epsilon_2$ such that
\begin{align}
M^{(4f)}_{QI}(\Theta,4\rho-4\psi) &= \epsilon_1(\Theta) \cos(4\rho - 4\psi + \phi_{QI})\\
M^{(4f)}_{UI}(\Theta,4\rho-4\psi) &= \epsilon_2(\Theta) \cos(4\rho - 4\psi + \phi_{UI}).
\end{align}
The dependence on $\Theta$ of those parameters is relatively large, and will be important to consider while including several detectors in the focal plane. For simplicity we dropped the $(4f)$ notation index for the phase. So otherwise specified, $\phi_{ij}$ refer to the phase of the $4f_{\rm HWP}$ terms for the rest of the paper. 

Let's notice at this point that the $M^{(0f)}$ matrix does not contribute to the polarization measurement.
Another important aspect to consider is that the values of the phases $\phi_{UI}$ and $\phi_{QI}$ depend on the reference frame used to define the angles ($\rho$ and $\psi$ in particular) which is defined for along an axis in the focal plane. In case other detectors at different locations in the focal plane are considered, we account for the phase changes as described later in Section~\ref{sec:multi-det}.  

Other Mueller matrix elements eventually produce mixing of $Q$ and $U$ components and/or alter the measurement of the intensity.

Given the required computation time with RCWA simulations, Mueller matrix coefficients are calculated for a very limited number of incident angles on the HWP and empirical relations are needed to predict the signal at any detector location on the focal plane. We use the parametrization of paper-I and assume the following model for the Mueller matrix:
\begin{equation}
M^{(if)}_{ij}(\Theta) = M^{(if)}_{ij}(0) +  \Big(M^{(if)}_{ij}(\Theta_{\rm ref}) - M^{(if)}_{ij}(0)\Big) \frac{f(\Theta)}{f(\Theta_{\rm ref})}
\label{eq:epsThetaMod}
\end{equation}
with the empirical function $f(\Theta)$: 
\begin{equation}
f(\Theta) = \sin^2(a\,\Theta),
\label{eq:modftheta}
\end{equation}
where $M(0)$ and $M(\Theta_{\rm ref})$ are calculated at normal incidence and for the reference angle $\Theta_{\rm ref}$, respectively. In the LFT 140-GHz band, the constant $a$ has been fixed to 0.078\,deg$^{-1}$ (for $\Theta$ in degree) to match the calculation of Mueller matrix IP coefficients for several incident angles in paper-I. We assume that the same transformation to the coefficients can be applied to any entry of the matrix $M_{XY}-M_{XY}^{\rm Ideal}$. This is a somewhat coarse approximation for some coefficients, but it is valid for both $4f_{\rm HWP}$ IP terms.
We also assume that the phases $\phi_{XY}$ are constant with respect to the angles $\Theta$, which is verified with a good approximation in paper-I. At the 140-GHz frequency band, the $4f_{\rm HWP}$ IP terms verify $M^{(4f)}_{XI}(0) =0$ and $\epsilon_1(\Theta_{\rm ref}) = \epsilon_2(\Theta_{\rm ref}) = 8.86\times 10^{-5}$ for $\Theta_{\rm ref}=10^\circ$. All the Mueller matrix amplitude coefficients and phases used as a test case in this paper are given in Tables~\ref{tab:MuellerCoeff140GHz} for 10$^\circ$ incident angle:
\begin{table}[htbp!]
 \centering
   {\setlength{\extrarowheight}{4pt}%
			\begin{tabular}{|c||c|c|c|c|}
				\hline
				& $M^{(0f)}_{xI}$ & $M^{(0f)}_{xQ}$ & $M^{(0f)}_{xU}$ & $M^{(0f)}_{xV}$\\
				\hline
				\hline
				  $M^{(0f)}_{Ix}$ & 0.961 & 8.83 $\times 10^{-5}$ & $-7.87 \times 10^{-6}$& 9.17 $\times 10^{-5}$ \\
				\hline
				$M^{(0f)}_{Qx}$ & 9.60 $\times 10^{-5}$ & $1.88 \times 10^{-4}$ &  4.87 $\times 10^{-4}$ & $-3.45 \times 10^{-3}$ \\
                \hline
				  $M^{(0f)}_{Ux}$ & 4.39 $\times 10^{-6}$ & $-4.63 \times 10^{-4}$ & 7.48 $\times 10^{-4}$ & 0.0212 \\
                \hline
				  $M^{(0f)}_{Vx}$ & $-9.34$ $\times 10^{-5}$ & $-1.29 \times 10^{-3}$ & $-0.0242$ & $-0.959$ \\
				\hline
			\end{tabular}
   }
\vspace{1em}

  {\setlength{\extrarowheight}{4pt}%
           \begin{tabular}{|c||c|c|c|c|}
				\hline
				& $M^{(2f)}_{xI}$ & $M^{(2f)}_{xQ}$ & $M^{(2f)}_{xU}$ & $M^{(2f)}_{xV}$\\
				\hline
				\hline
				  $M^{(2f)}_{Ix}$ & 4.89 $\times 10^{-6}$ & 5.15 $\times 10^{-4}$ & 5.16 $\times 10^{-4}$& 2.64 $\times 10^{-5}$ \\
				\hline
				$M^{(2f)}_{Qx}$ & 5.43 $\times 10^{-4}$ & 3.10 $\times 10^{-3}$ & 3.28 $\times 10^{-3}$ & 0.0231 \\
                \hline
				  $M^{(2f)}_{Ux}$ & 5.42 $\times 10^{-4}$ & 2.96 $\times 10^{-3}$ & 3.24 $\times 10^{-3}$ & 0.0230 \\
                \hline
				  $M^{(2f)}_{Vx}$ & 4.61 $\times 10^{-5}$ & 0.0231 & 0.0231 & 1.04 $\times 10^{-3}$ \\
				\hline
			\end{tabular}
   }
\vspace{1em}

  {\setlength{\extrarowheight}{4pt}%
           \begin{tabular}{|c||c|c|c|c|}
				\hline
				& $M^{(4f)}_{xI}$ & $M^{(4f)}_{xQ}$ & $M^{(4f)}_{xU}$ & $M^{(4f)}_{xV}$\\
				\hline
				\hline
				  $M^{(4f)}_{Ix}$ & 1.09 $\times 10^{-7}$ & 9.26 $\times 10^{-5}$ &  9.25 $\times 10^{-5}$ & 1.97 $\times 10^{-6}$ \\
				\hline
				$M^{(4f)}_{Qx}$ & $\boldsymbol{8.86 \times 10^{-5}}$ & 0.959 & 0.959 & 0.0241 \\
                \hline
				  $M^{(4f)}_{Ux}$ & $\boldsymbol{8.86 \times 10^{-5}}$ & 0.959 & 0.959 & 0.0241 \\
                \hline
				  $M^{(4f)}_{Vx}$ & 1.58 $\times 10^{-6}$ & 0.0214 & 0.0214 & 5.55 $\times 10^{-4}$ \\
				\hline
			\end{tabular}
   }
   \vspace{1em}

  {\setlength{\extrarowheight}{4pt}%
             \begin{tabular}{|c||c|c|c|c|}
				\hline
				& $\phi^{(2f)}_{xI}$ & $\phi^{(2f)}_{xQ}$ & $\phi^{(2f)}_{xU}$ & $\phi^{(2f)}_{xV}$\\
				\hline
				\hline
				  $\phi^{(2f)}_{Ix}$ & $-2.32$ & $-0.49$ & $-2.06$ & 1.20 \\
				\hline
				$\phi^{(2f)}_{Qx}$ & 2.86 & $-0.25$ & $-2.00$ & 0.12 \\
                \hline
				  $\phi^{(2f)}_{Ux}$ & 1.29 & $-2.01$ & 2.54 & $-1.45$ \\
                \hline
				  $\phi^{(2f)}_{Vx}$ & $-1.94$ & 2.30 & 0.73 & 1.16 \\
				\hline
			\end{tabular}
   }
   \vspace{1em}

   {\setlength{\extrarowheight}{4pt}%
             \begin{tabular}{|c||c|c|c|c|}
				\hline
				& $\phi^{(4f)}_{xI}$ & $\phi^{(4f)}_{xQ}$ & $\phi^{(4f)}_{xU}$ & $\phi^{(4f)}_{xV}$\\
				\hline
				\hline
				  $\phi^{(4f)}_{Ix}$ & $-0.84$  & $-0.04$ & $-1.61$ & $-1.47$ \\
				\hline
				$\phi^{(4f)}_{Qx}$ & 0.14 & $-0.00061$ & $-0.00056 - \pi/2$ & $-1.63$  \\
                \hline
				  $\phi^{(4f)}_{Ux}$ & $-1.43$& $-0.00070 - \pi/2$& $\pi$ $-0.00065$ & 3.08 \\
                \hline
				  $\phi^{(4f)}_{Vx}$ & 1.91 & 1.72 & 0.15 & 0.09  \\
				\hline
			\end{tabular}
   }
\caption{
\label{tab:MuellerCoeff140GHz} Band averaged HWP Mueller matrix amplitude coefficients and phases used in our study of the 140-GHz band for a 10$^\circ$ incident angle and the 0th, 2nd and 4th harmonics of the rotation frequency as described in Equation~\ref{eq:epsThetaMod}. The matrix is expressed in focal plane coordinates and is calculated with RCWA simulation for a nine-layer HWP used for the \LB Low-Frequency Telescope (see~\cite{Imada18}, \cite{Kunimoto19}). The coefficients indicated in bold are at the origin of the instrumental polarization and they are inducing the largest systematics at low multipole included in this formalism.
}
\end{table}

Let's notice that a $\sin^4(b\,\Theta)$ term of the empirical function $f$ has been fitted to the data and has a small but significant contribution for detectors at the edge of the focal plane. We chose to neglect this term which has a small impact on the final contribution to the spectra for simplicity. This paper provides a framework for future analysis and more detailed models can be used as input. Accurate studies applied on real data will require modeling based on ground calibration.

\subsection{Instrument Configuration}

We use the \LB nominal instrumental configuration described in~\cite{PTEP}. We assume a HWP continuously rotating at 46\,rpm, and several different configurations of locations of 72 detector pairs in the focal plane. Detectors within each pair are sensitive to orthogonal polarization, and there is a 45 degree shift of polarization angle between two adjacent pairs. We use standard scanning strategy for a satellite mission as adopted by \LB\cite{LB2021} with a rotation axis precessing around the Sun-Earth-L2 axis, and with parameters $\alpha=45^{\circ}$ for the precession angle, $\beta=50^{\circ}$ for the observation angle with respect to the rotation axis, $T_{\rm prec}=192.348$\,min the precession period and $T_{\rm rot}=20$\,min the rotation period. The ratio of periods has been optimized to reduce resonances in the scanning strategy as described in~\cite{BPMM}.
Data simulations are performed for the \LB LFT frequency band centered at 140\,GHz~\cite{Sekimoto21} and include simulated sky emission at that frequency as detailed in the next part. We assume an acquisition frequency of 19\,Hz. Noise is added optionally depending on the considered case, in particular for the evaluation of the efficiency of the subtraction of the IP leakage. It is not necessary to add noise to the timestreams for evaluating the effect of HWP imperfection on the signal without correction since our map-making method is linear. We use the nominal noise of \LB detectors described in~\cite{PTEP}. 

\subsection{Sky Model}

The model of Stokes parameters $I,Q,U$ of the sky emission at a given position ${\vec{r}}$ can be written as a sum of contribution from different sky components:
\begin{align}
 I(\vec{r}) = I_0 + D + \sum_c I_c ;\,\, Q = \sum_c Q_c ;\,\, U = \sum_c U_c;\,\, V=0,
\end{align}
with $I_0$ the CMB monopole signal at $T=2.735\,$K, $D$ the solar dipole, $I_c, Q_c, U_c$ the Stokes parameters of component $c$. We explicitly write the monopole signal since it is expected, as other components, to leak into polarization signal estimation due to instrumental polarization. We include as components, both in polarization and intensity, the CMB, the Galactic dust and the Galactic synchrotron emission. We also include the unpolarized emission from free-free and spinning dust components. We use the pySM model (D0S0) to describe the components with fixed component spectra (no spatial variations). This has little importance for this analysis since we evaluate the IP leakage in a single frequency channel. Finally we assume no $V$ polarization from the CMB or foregrounds. Although foreground might emit some small amount of $V$ polarization, the only way to observe those in through HWP imperfections.

Although the monopole signal contribution can be filtered with minimal impact on the final maps for an ideal instrument, as leakage would appear at four times the HWP rotation frequency in the modulated timestreams, the coupling of this large modulated signal with other systematic effects such as nonlinearities or time-varying gains could induce broader spectrum which would project to the polarization maps. Consequently, even if time varying loading or nonlinearities are not included in this study, we keep this term in the expressions to allow extensions of the formalism (in particular for the correction section). Similar arguments apply to the dipolar signal since a template could be fitted and removed from the data, but the main difference is that it be used as calibrator for the IP coefficients. This will be largely discussed in the last section of the paper.

\section{Predictions of the Effect on the Reconstructed Maps}
\label{sec:Prediction}

In this section we evaluate the bias on the spectra induced by the imperfections of the HWP, focusing mostly on instrumental polarization which is the dominant effect due to the large amplitude of the intensity signal. We also provide a link between the coefficients $\epsilon_1$ and $\epsilon_2$ for a future satellite mission such as \LB and the induced bias on the tensor-to-scalar ratio $r$.

\subsection{Simulations and Results}

\subsubsection{Data Simulations}

We start from simulations of the sky at 140\,GHz containing band integrated components assuming a simple top-hat band-pass filter. Input maps for simulations are derived from pySM~\cite{PSM17} and are represented using HEALPix\footnote{{\it http://healpix.sourceforge.net}~\cite{Healpix}}. We then generate pure signal timestreams by simulating the scanning of the sky with a Gaussian beam of 23.7\,arcmin FWHM and an imperfect HWP by implementing the model~\ref{eq:modelobs}.
We simulate when necessary the timestreams for the 144 detectors of the band (72 pairs). We did not integrate the signal over the sampling period but simply extracted the signal from the maps. A more realistic treatment should not affect the results detailed in this paper and should only have an effect on the polarization efficiency. When relevant, we add noise including a 1/f component with an arbitrary (pessimistic) knee frequency $f_{\rm knee}=50$\,mHz. For this analysis we make two important approximations as was previously stated. First we assume that the main beam extent for each detector is much smaller than the scale of variation of $M({\Theta})$, i.e. that there is no beam shape changes introduced by HWP imperfections which might then depend on the HWP angle $\rho$. Second, we neglect color effects due to different spectra of component integrated on a broad band leading to slightly different Mueller matrix coefficients depending on the sky component (this results in a few percent difference between CMB and Galactic dust for the considered band as calculated in paper-I). This has been shown to be small, especially since the CMB dipole is the main source of the IP spurious signal and after applying a mask covering the Galactic plane. We assume a CMB spectrum for the calculation of band averaged coefficients.

\subsubsection{Analysis}

The Stokes parameter maps $\widehat{I}$, $\widehat{Q}$, $\widehat{U}$ are reconstructed assuming an ideal HWP (ignoring the imperfections which are introduced in simulations) using as input the simulated TOIs. The effect of imperfections is evaluated from residual maps, after subtracting the true input sky maps.

With an ideal HWP, the data model reduces to:
\begin{equation}
s_{\rm ideal} = I + \cos (4\rho - 2(\psi + \psi_0))\,Q + \sin (4\rho - 2(\psi + \psi_0))\,U + n,
\end{equation}
and define the matrix $W$ such that for the collection of all data samples $\vecs_{\rm ideal}$ we have:
\begin{equation}
\vecs_{\rm ideal} = W \vecm + \vecn,
\end{equation}
with $\vecm=\{\vecI,\vecQ,\vecU\}$ the collection of Stokes parameter maps, and the matrix $W=PM_{\rm ideal}$ is the product of the detector pointing matrix $P$ and an ideal Mueller matrix. Each element of $M$ relates time $t$ and pixel $p$ such that $W_{pt} = (1; \cos (4\rho - 2(\psi + \psi_0)) ; \sin (4\rho - 2(\psi + \psi_0)))$.
The maximum likelihood solution for the maps under the assumption of Gaussian noise is:
\begin{equation}
 \widehat{\vecm} = ({W}^t N^{-1} W)^{-1} {W}^t N^{-1} \vecs
\end{equation}
where $\widehat \vecm$ are the recovered Stokes parameter maps, $N=<\vecn^t\vecn>$ is the noise covariance matrix in time domain which is calculated from the input noise power spectrum $P(f)$ assuming stationarity: $N^{-1}=F^{-1} P^{-1} F$, with $F$ the Fourier operator and with the model: $P(f) = P_0 [(f/f_{\rm knee})^{-1} + 1]$ and $f_{\rm knee} = 0.05$\,Hz. We use the SANEPIC GLS code~\cite{Patanchon08} for solving the map-making equation. The solution for polarization maps with a simple projector (i.e. assuming white noise and setting $N$ to the identity matrix) should be nearly optimal for a fast rotating HWP, however, the reconstruction of the temperature is significantly improved when using the proper noise power spectra as compared to using simpler methods. Also, the computation time of the solution with SANEPIC is relatively short with moderate computer resources when performed on the single detector basis. The multi-detector case can be dealt with in a nearly optimal way by combining recovered single detector maps. The computation of one set of Stokes parameter maps takes 20 minutes with 18 processors for one year observation with a single detector.

\begin{figure}
  \centering
  \includegraphics[width=0.495\columnwidth]{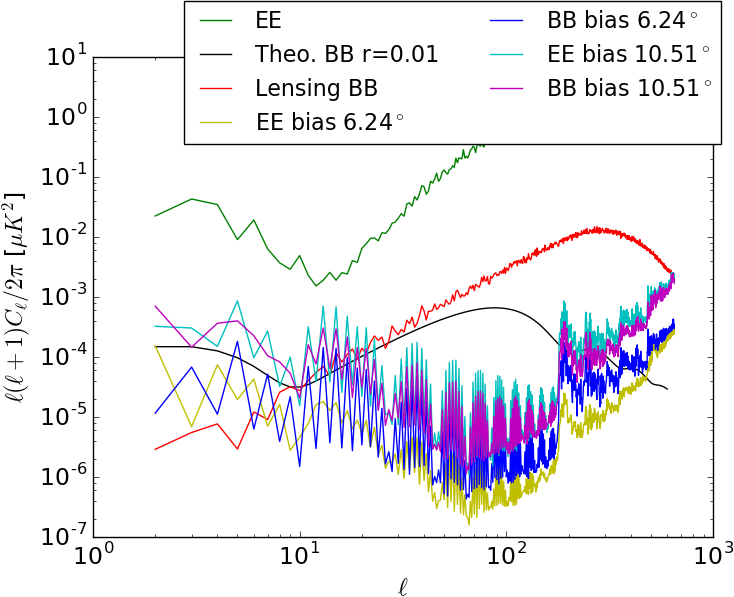}
  \includegraphics[width=0.495\columnwidth]{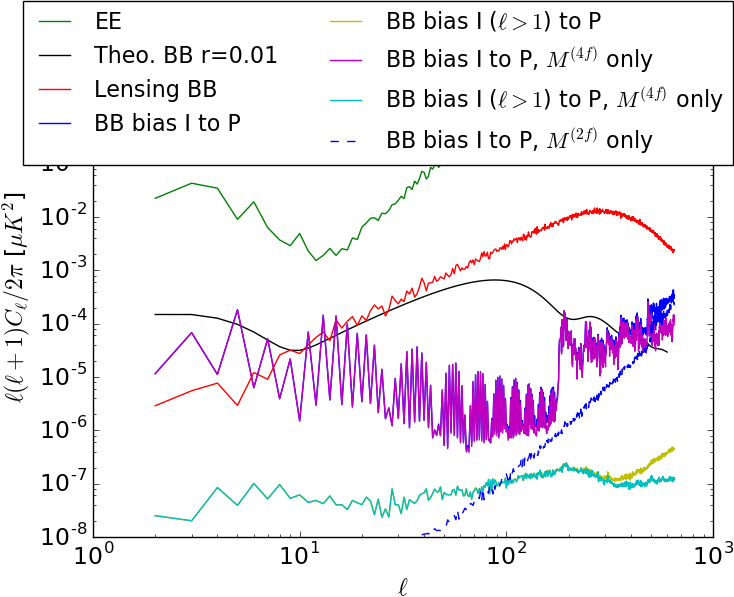}
  \caption{Left: Polarization $E$- and $B$-mode power spectra of the residual maps from HWP imperfections for two detectors at different locations in the focal plane using the Mueller matrix coefficients from RCWA simulations (the incident angle $\Theta$ is indicated in the legend). Input TOI simulations include CMB anisotropies with cosmological dipole, Galactic dust, synchrotron and free-free emission. The instrumental polarization is the dominant source of the bias of the polarization power spectra. The amplitudes of the 4-$f_{\rm HWP}$ Mueller matrix IP coefficients $\epsilon_1$ ($\epsilon_2 = \epsilon_1$) for the two detectors are $3.9 \times 10^{-5}$ and $9.6 \times 10^{-5}$. Right: Contributions to the leakage power spectra from the dipole and from the anisotropies (including CMB and foregrounds) separately including the contribution of IP only, and for the detector with an incident angle $\Theta = 6.24^\circ$. For each case the leakage power spectra are estimated with or without the contributions of $M^{(2f)}$ terms. For each case, recovered polarization maps have been recalibrated before making the difference with the true input in order to remove the effect of non unitary diagonal Mueller coefficients (assuming then that those coefficients could be marginalized over during the in-flight calibration procedure). Apart from $\ell$ to $\ell$ oscillations, most of the differences between the $EE$ and $BB$ power spectra are due to $E$ to $B$ leakage caused by imperfect $M^{(4f)}_{QQ/QU/UQ/UU}$. An apodised mask of the Galactic plane has been applied to the polarization maps leaving 50\,\% of the sky for the computation of power spectra. The spectra are corrected from partial coverage by dividing by a simple $f_{\rm sky}$ fraction. The dash-blue curve shows the spectrum of the IP leakage from $M^{(2f)}$. The curve not shown at low $\ell$ flattens and reaches $\approx 10^{-11}\,\mu$K$^2$.}
    \label{SpSigleak}
\end{figure}

Figure~\ref{SpSigleak} shows the bias induced on the $EE$ and $BB$ power spectra using data simulations and assuming two different detectors located at $\Theta=6.24^\circ$ and $\Theta = 10.51^\circ$ from the HWP rotation axis. This leads to different amount of IP for the two detectors of $\epsilon_1 = \epsilon_2 = 3.9\times 10^{-5}$ and $9.6\times 10^{-5}$. The coordinate azimuthal angles around the HWP rotation axis are not identical on purpose for the two detectors due to two different chosen locations in the focal plane, leading to different phases of the IP and then to small changes in the shape of the power spectra.
To compute the power spectra displayed in the Figure, reconstructed noiseless residual maps are recalibrated by multiplying by a factor $\alpha$ very close to unity such that the hit weighted difference between the recovered maps and the input (true) maps is minimized. More specifically we calculate ${\rm min}|_{\alpha_1} \{\sum_p N_p (\widehat Q_p - \frac{1}{\alpha_1} Q_p)\}$ and ${\rm min}|_{\alpha_2} \{\sum_p N_p (\widehat U_p - \frac{1}{\alpha_2} U_p)\}$ and take the averaged parameter $\alpha = \frac{1}{2}(\alpha_1+\alpha_2)$. We then calculate the power spectrum from $\delta \widehat Q = \alpha \widehat Q - Q$ and $\delta \widehat U = \alpha \widehat U - U$. This operation, even if it can not be used in the same way with real data, allows to isolate the contribution of instrumental polarization, as well as $Q/U$ mixing effect for the evaluation of the bias without mixing with other artefacts affecting calibration and polarization efficiency. Indeed, other Mueller matrix parameters such as those associated to the amplitude of the sub-matrix elements $M^{(4f)}_{ij}$ with $i,j=Q,U$ have the effect of reducing the polarization maps by a global factor. 
The applied procedure could in principle bias the estimation of the IP signal but since the fitted $Q$ and $U$ Stokes parameter maps are not correlated with the residual IP, this effect is negligible. This operation is then close to a perfect re-calibration of the maps.

Finally, the power spectra are computed after masking 50 percent of the sky over the Galactic plane using the publicly available apodised mask provided by \Planck\ for the polarization analysis\footnote{We use the mask {\it COM\_Mask\_Likelihood-polarization-143\_2048\_R2.00.fits}
available at \\{\it https://pla.esac.esa.int/\#maps}.}. The power spectra are rescaled by a simple $f_{\rm sky}$ factor.

For the detector with instrumental polarization parameter at 4$f_{\rm HWP}$ $3.9\times 10^{-5}$ we observe a contribution of the leakage to the maps with an amplitude of more than 10 times the $BB$ lensing signal at large scale $\ell<10$. The residual is for most caused by the cosmological (or solar) dipole. We verify that the leakage spectrum is multiplied by a factor of approximately 8 at low multipoles when coefficients are multiplied by 2.46, although the phase difference makes the comparison between the two spectra approximate. This is in agreement with an expected quadratic scaling of the power spectrum with the amount of IP.

We also evaluate the separate contributions of the effect of the $2f_{\rm HWP}$ and $4f_{\rm HWP}$ components of the Mueller imperfection matrix. For the considered detector in the Figure (right side), the amplitude of the 2$f_{\rm HWP}$ component is $2.39\times 10^{-4}$. This component induces a leakage with a slightly raising $BB$ power spectrum. The overall effect is a relatively small, and it which would vanish if the HWP would rotate in each pixel by an exact angle of $n_p \pi$ with $n_p$ an integer at the limit of infinite sampling. The 2$f_{\rm HWP}$ component leads to a negligible bias on the tensor-to-scalar ratio, for the parameters used for in our modeling, and is moreover reduced by a factor $\frac{1}{N_p}$, with $N_p$ the number of data point in a given pixel (this has been checked with our simulations). It is further decreased when combining several detectors. We have verified with simulations that the parameters of the $M^{(0f)}$ matrix have no influence on the polarization results.

We have also run simulations with each sky component, the temperature anisotropies, the polarization, and cosmological dipole taken separately and we computed the residual power spectra on $E$ and $B$ modes. The CMB anisotropies contribute to a small and negligible fraction of the total IP as expected.

The shape of the residual power spectra showing power of the systematic effect at all scales is well explained, in particular considering the effect of the dipole, as the leakage IP map is the result of the modulation of the intensity signal with the spin-2 scanning cross-linking parameter~\footnote{The scanning cross linking parameter are defined as $<cos2\psi>$ and $<sin2\psi>$ with $\psi$ the focal plane orientation angle on the sky.}, containing power at all multipoles after projection on the Stokes parameter maps. We will develop analytical expressions in the next section~(\ref{sub:analytical}). Provided that the dipole orientation and amplitude are known, the effect on $Q$ and $U$ maps can be predicted analytically with respect to a limited number of Mueller matrix parameters.

The $EE$ and $BB$ power spectra of the residuals for individual detectors have similar shapes and amplitudes, except for large $\ell$ to $\ell$ oscillations with phases dependent on the detector location in the focal plane and on the phase of the imperfections in the Mueller matrix. There is a small but significant difference between the two power spectra at intermediate scales, for $\ell \approx 100$ due to $Q$ and $U$ mixing induced by non-ideal elements of the Mueller matrix $M_{QQ}$, $M_{QU}$, $M_{UQ}$ and $M_{UU}$, which leads to $E$ to $B$ leakage (and to a lesser extent to $B$ to $E$ leakage).

\begin{figure}
  \centering
  \includegraphics[width=0.49\columnwidth]{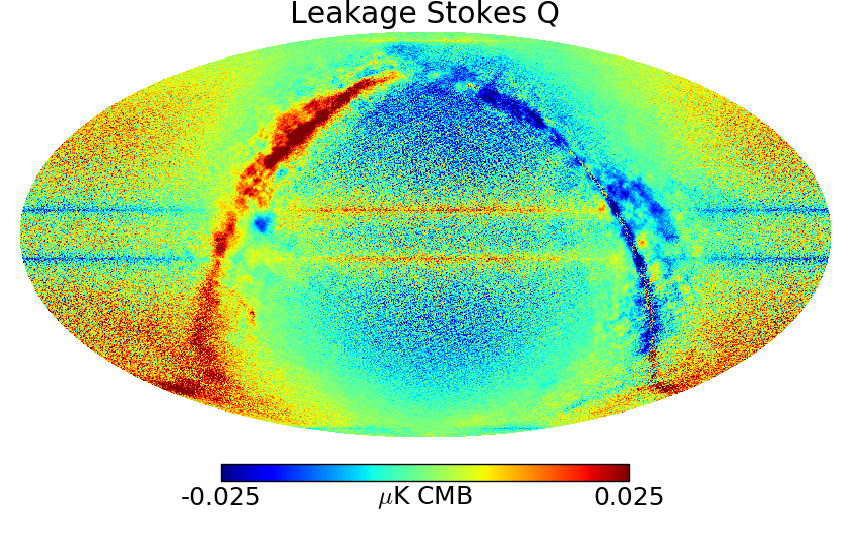}
  \includegraphics[width=0.49\columnwidth]{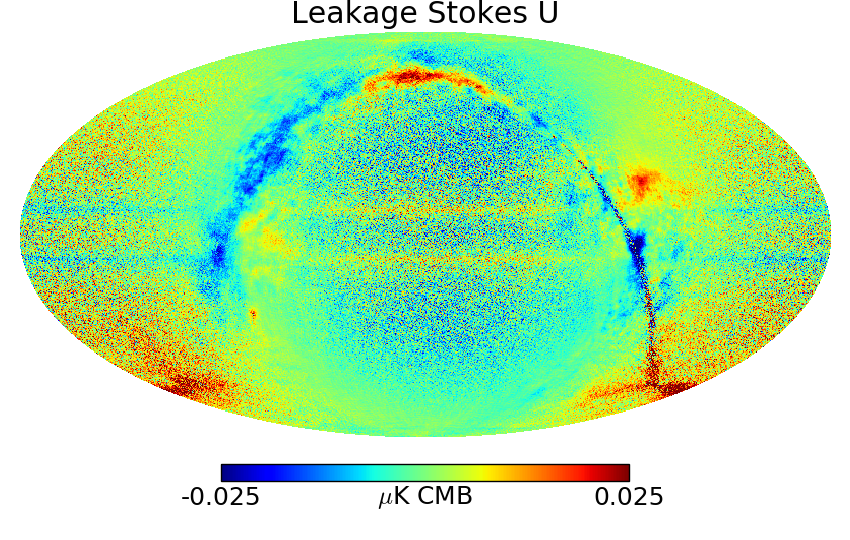}
  \includegraphics[width=0.49\columnwidth]{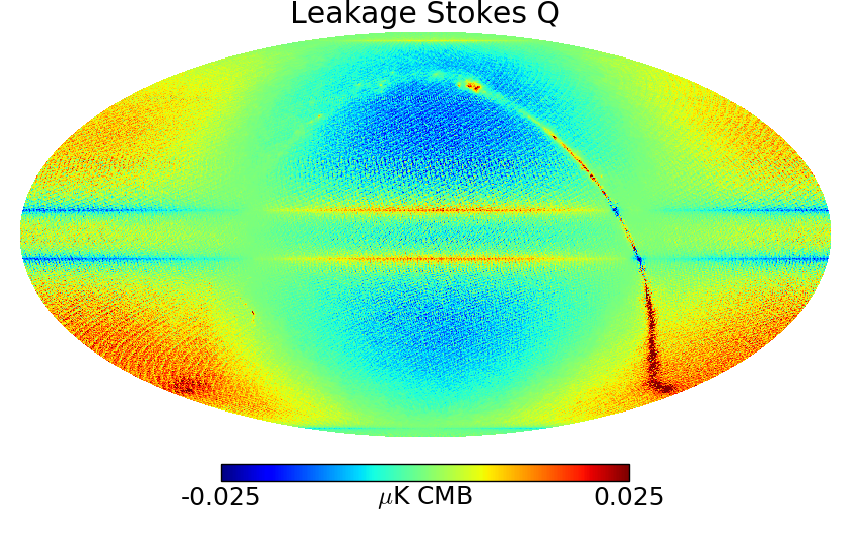}
  \includegraphics[width=0.49\columnwidth]{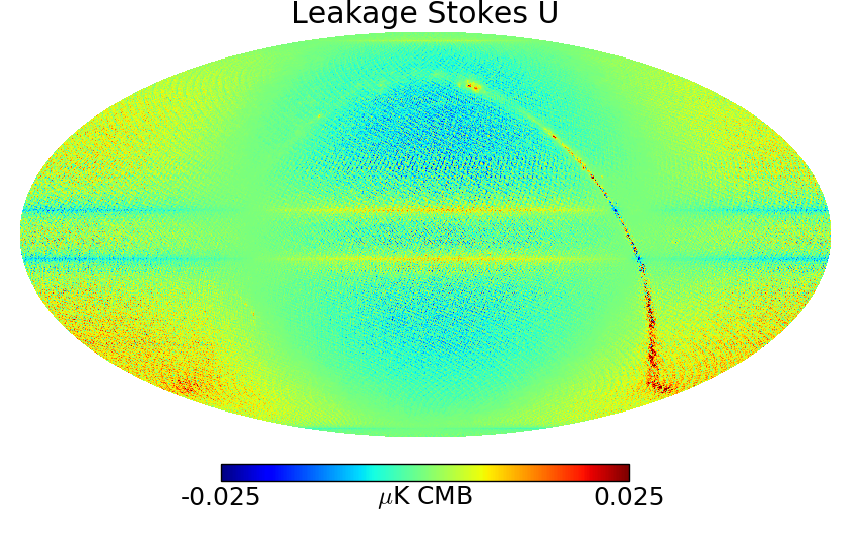}
  \caption{Residual Stokes parameter maps $\delta \widehat Q$ (left) and $\delta \widehat U$ (right) in ecliptic coordinates for one detector observing at $\Theta=3.96^\circ$ with an assumed amount of IP at $4f_{\rm HWP}$ of $\epsilon_1 = \epsilon_2 = 1.65\times 10^{-5}$. Input data simulations include cosmological dipole, CMB anisotropies and Galactic foregrounds intensity and polarization and the Mueller matrix of the HWP from RCWA simulations. The maps are obtained using SANEPIC map-making assuming an ideal HWP. The $2f_{\rm HWP}$ IP amplitude is $1.01\times 10^{-4}$. The residual maps are obtained after removing the input polarization maps. They include in the first row the residual effects from the 2 and $4f_{\rm HWP}$ components of the Mueller matrix, and in the second row the residuals with the $4f_{\rm HWP}$ IP only (no effect of polarization $E/B$ mixing and IP at $2f_{\rm HWP}$ included).}
    \label{Leakmaps}
\end{figure}

Figure~\ref{Leakmaps} shows the residual maps $\delta \widehat Q$ and $\delta \widehat U$ for one detector located at $\Theta=3.96^\circ$. The residual effect from the full Mueller matrix imperfections is shown on the top, while the contribution of the IP at $4f_{\rm HWP}$ only is shown at the bottom (for this detector $\epsilon_1 = 1.65\times 10^{-5}$). The effect of the dipole is clearly visible in the maps. The modulation of the projected temperature signal by the scanning cross-linking parameters can also be observed with the horizontal lines where the scanning angle uniformity measurement is the poorest.
We can observed the effect of the $2f_{\rm HWP}$ component of the matrix inducing a low amplitude high spatial frequency component in the maps (see discussion above), by comparing the maps with and without this term. Finally the effect of $Q$ and $U$ mixing can be observed along the Galactic plane (where the polarization signal is the strongest) in the residual maps, where an excess signal is visible in addition to the IP leakage. This effect does not depend on the orientation of the focal plane and so is not modulated after projection onto the maps by the cross-linking parameters. This can also be seen in the residual maps. 

When several detectors are used and when the recovered polarization maps are averaged, we expect efficient cancellations of the IP leakage. In particular we will show that two detectors located at the same $\Theta$ and 90$^\circ$ apart around the HWP rotation axis have opposite leakage effects on timestreams. This reduces significantly the impact of instrumental polarization on the final combined maps. This can also be used to constrain the instrumental polarization parameters. Before addressing the multi--detector case, we focus on the analytical calculation of the effect of the IP. 

\subsection{Analytical Calculations}
\label{sub:analytical}

In order to predict the impact of the leakage to polarization let's first compute the bias induced by the instrumental polarization to the measured signal at any given time. We write  $\delta M$ as the excess Mueller matrix with respect to the ideal case considering only the instrumental polarization at 4$f_{\rm HWP}$, taking only $\delta M_{QI} =\epsilon_1\cos (4\rho - 4\psi + \phi_{QI})$ and $\delta M_{UI} =\epsilon_2\cos(4\rho - 4\psi + \phi_{UI})$ coefficients as entries (other entries are set to 0) and assuming arbitrary phases $\phi_{QI}$ and $\phi_{UI}$ at this point:
\begin{eqnarray}
\delta s &=& \Big(1 \,\,\cos 2\psi_0 \,\,\sin 2\psi_0 \Big)\,
\delta M
R(2\psi)
\begin{pmatrix}
I \\ Q \\ U
\end{pmatrix},\\
& = & [\epsilon_1 \cos(4\rho - 4\psi + \phi_{QI}) \cos 2\psi_0 + \epsilon_2 \cos(4\rho - 4\psi + \phi_{UI}) \sin 2\psi_0]\,I,
\label{eq:delts}
\end{eqnarray}
where we are ignoring the Stokes parameter $V$ since we do not consider it in our sky model, and are not sensitive to it with a polarizer. The Mueller matrices are then 3 by 3 matrices from here and we keep the same notations as in previous sections. 

Because the leakage term is modulated with a phase $4\rho - 4\psi$ and the projection term with $4\rho - 2(\psi + \psi_0)$, we can understand at this point that a high degree of cross--linking of the scanning (i.e. with many different values of $\psi$ angles in each pixel) will lead to less leakage in the final maps.
To verify that,
let's propagate the leakage component through the map--making equation assuming white noise only for simplification:
\begin{equation}
	\delta \vecm = (W^t W)^{-1} W^t\,\delta \vecs,
	\label{eq:mmdelts}
\end{equation}
with the $W$ matrix defined above. Let's define $\chi=2\rho - (\psi + \psi_0)$, writing this solution pixel by pixel, which is optimal since white noise is assumed, gives:
\begin{eqnarray}
\begin{pmatrix}
\phantom{\bigg|}\delta I(p)\\
\phantom{\bigg|}\delta Q(p)\\
\phantom{\bigg|}\delta U(p)
\end{pmatrix}
=
\begin{pmatrix}
\phantom{\bigg|}1 &
\phantom{\bigg|}\left< \cos 2\chi_j \right> &
\phantom{\bigg|}\left< \sin 2\chi_j \right> \\
\phantom{\bigg|}\left< \cos 2\chi_j \right> &
\phantom{\bigg|}\dfrac{1+\left< \cos 4\chi_j \right> }{2} &
\phantom{\bigg|}\dfrac{\left< \sin 4\chi_j \right> }{2} \\
\phantom{\bigg|}\left< \sin 2\chi_j \right> &
\phantom{\bigg|} \dfrac{ \left< \sin 4\chi_j \right> }{2}  &
\phantom{\bigg|} \dfrac{1- \left< \cos 4\chi_j \right> }{2}
\end{pmatrix}^{-1}
\times
\begin{pmatrix}
  \phantom{\bigg|}\left< \delta s_j\right>\\
  \phantom{\bigg|}\left< \cos 2\chi_j \,\delta s_j\right>\\
  \phantom{\bigg|}\left< \sin 2\chi_j \,\delta s_j\right>
\end{pmatrix}
\label{ProjEq}
\end{eqnarray}
where $j$ denotes all samples falling in pixel $p$, and $\left< . \right>$ the average of a quantity
over all data samples $j$. Because the angle $\chi_j$ spans many different values in the interval $[0,2\pi]$ in each pixel, the matrix $W^tW$ is nearly diagonal since $\left< \cos 2\chi_j \right> << 1$, $\left< \cos 4\chi_j \right> << 1$, $\left< \sin 2\chi_j \right> << 1$, and $\left< \sin 4\chi_j \right> << 1$.
Then, the leakage maps are simply after replacing $\delta s$ by its expression~\ref{eq:delts} in Eq.~\ref{eq:mmdelts}:
\begin{align}
&\delta I \approx 0\\
&\delta Q \approx \Big[\epsilon_1 \left< \cos (2\psi - 2\psi_0 - \phi_{QI})\right> \cos 2\psi_0\, +\epsilon_2 \left< \cos (2\psi - 2\psi_0 - \phi_{UI})\right> \sin 2\psi_0\Big]\,I\\
&\delta U \approx \Big[\epsilon_1 \left<  \sin(2\psi - 2\psi_0 - \phi_{QI})\right> \cos 2\psi_0 + \epsilon_2\left<\sin(2\psi - 2\psi_0 - \phi_{UI})\right> \sin 2\psi_0\Big]\,I,
\label{Prediction}
\end{align}
where we have neglected all the terms for which the phases depend on the rotation angle of the HWP $\rho$.\footnote{We expect that those terms average to zero at large scales even for HWP rotation frequencies as low as 1\,rpm, for a scanning strategy with a rotation period of 20 minutes.}
The leakage maps are then the result of the product of the intensity map with a linear combinations of the second order cross-linking parameters of the scanning direction.
This is somewhat similar behaviour than the band-pass mismatch effect while combining different detectors without HWP as described in~\cite{BPMM}. Let's re-express the leakage components slightly differently:
\begin{align}
&\delta Q =  \Big[\frac{\epsilon_1}{2} \left< \cos (2\psi - \phi_{QI}) + \cos (2\psi - 4\psi_0 - \phi_{QI})\right> & \\ &   + \frac{\epsilon_2}{2} \left< \sin (2\psi - \phi_{UI}) - \sin (2\psi - 4\psi_0 - \phi_{UI})\right>\Big]\,I\\
&\delta U =  \Big[\frac{\epsilon_1}{2} \left< \sin (2\psi - \phi_{QI}) + \sin (2\psi - 4\psi_0 - \phi_{QI})\right> & \\ &  +  \frac{\epsilon_2}{2} \left< \cos (2\psi - 4\psi_0 - \phi_{UI}) - \cos(2\psi - \phi_{UI})\right>   \Big]\,I,
\end{align}
It's worth noticing that $\delta Q$ and $\delta U$ are independent of the polarizer angle $\psi_0$ if $\epsilon_1 = \epsilon_2$ and $\phi_{QI} = \phi_{UI} + \frac{\pi}{2}$. The leakage maps expressions simply become in that case:
\begin{align}
&\delta Q = \epsilon_1 \left< \cos (2\psi - \phi_{QI}) \right>\,I\label{lQ}\\
&\delta U = \epsilon_1 \left< \sin (2\psi - \phi_{QI}) \right>\,I.\label{lU}
\end{align}
Very importantly, RCWA simulations indicate that the conditions $\epsilon_1 = \epsilon_2$ and $\phi_{QI} = \phi_{UI} + \frac{\pi}{2}$ are satisfied within numerical uncertainties of RCWA simulations ($10^{-3}$ relative errors on parameters). The physical reason explaining why those amplitude and phase relations hold to a good accuracy is explained in Appendix~\ref{Ap:A}. In this particular case the HWP behaves like a fixed polarizer in the frame of the focal plane, and in this way the leakage on $\delta Q$ and $\delta U$ is independent of the detector polarizer orientations. We then find that the HWP IP systematics are tightly coupled with the spin-2 scanning cross linking parameters. This behaviour is observed for other systematics as described in~\cite{Wallis17} and \cite{McCallum21}.

The previous result is very important because it implies that there is no cancellation of the IP leakage by combining data from two detectors with perpendicular polarization within the same pair, with instead unfortunately leakage adding up coherently between different detectors at the same location. With the phase difference of the two terms in the Mueller matrix, the HWP behaves like a polarizer, with however a very low efficiency which depends on the the incident angle $\Theta$. 

In case of coloured noise, we do not expect the conclusions to change significantly. We expect only minor changes in the leakage maps since the GLS map-making solution (neglecting IP polarization) is almost equivalent to a simple projector in presence of a fast rotating HWP for the Stokes $Q$ and $U$ parameters. 

\begin{figure}
  \centering
  \includegraphics[width=0.49\columnwidth]{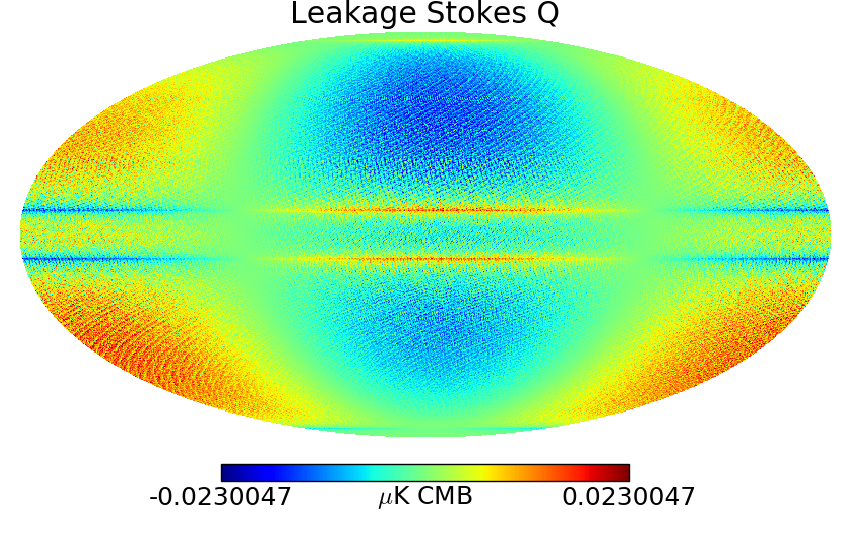}
  \includegraphics[width=0.49\columnwidth]{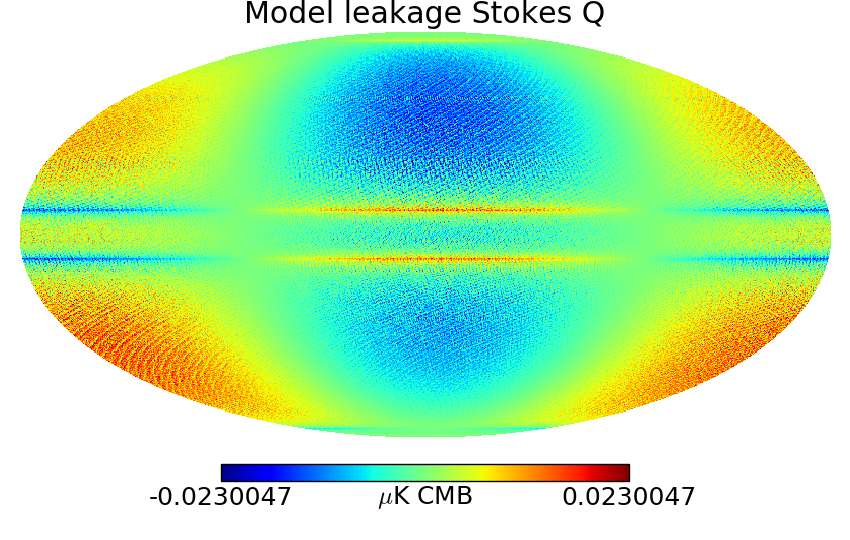}
  \includegraphics[width=0.49\columnwidth]{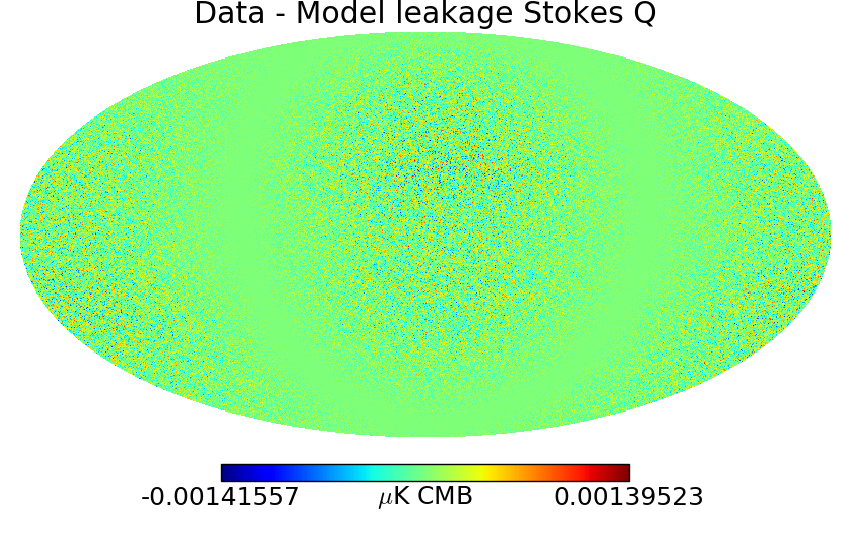}
  \includegraphics[width=0.49\columnwidth]{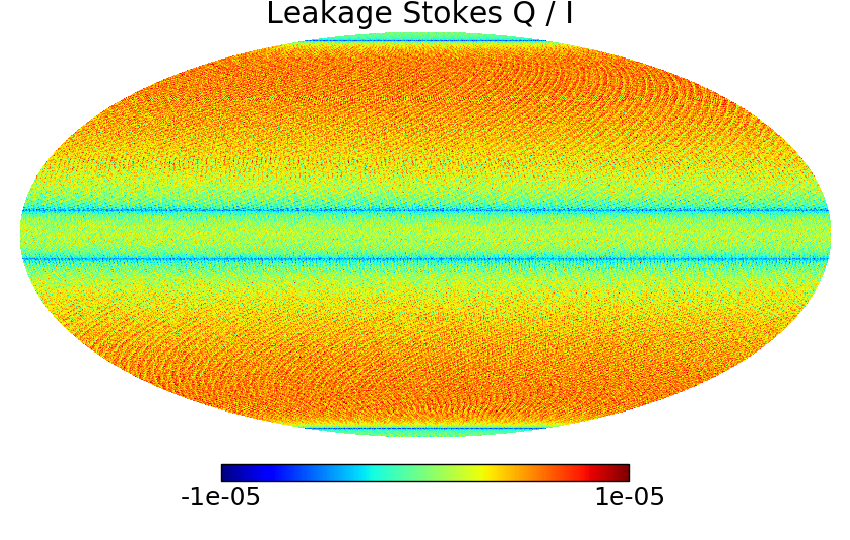}
  \caption{Top left: Leakage $\delta \widehat Q$ map obtained with dipole only simulations and imperfect Mueller matrix $M^{(4f)}$ for one detector observing at $\Theta=3.96^\circ$. Top right: maps of $\left< \cos (2\psi - \phi_{QI}) \right> I$ after adjusting the phase $\phi_{QI}$ to account for the detector location in the focal plane and taking the input dipole map as intensity $I$. Bottom left: difference between the leakage $\delta Q$ map and the model. Bottom right: map of $\frac{\delta Q}{I}$.}
    \label{QOverI}
\end{figure}

We verify the tight relation between the leakage maps and the spin-2 scanning cross-linking parameters $\left< \cos (2\psi - \phi_{QI}) \right>$ and $\left< \sin (2\psi - \phi_{QI}) \right>$ using our simulations including the dipole only as input signal and assuming only the imperfect Mueller matrix $M^{(4f)}$, i.e setting the matrix $M^{(2f)}=0$. We compare in Figure~\ref{QOverI} the reconstructed polarization $\delta Q$ map using SANEPIC (and assuming an ideal HWP for the reconstruction) with the map of the cross--linking parameter $\left< \cos (2\psi - \phi_{QI}) \right> I$, after adjusting the phase $\phi_{QI}$ for the same detector location in the focal plane as chosen for Figure~\ref{Leakmaps}, and taking the input dipole map as intensity $I$. We observe a very tight correlation of the two quantities confirming the relevance of the approximations. The difference between the two quantities, which is shown in the Figure, is roughly at the level of 2\,\% of the residual map. We observe pixel scale variations that can be explained by random variations around the mean of the scanning cross--linking parameters.

The analytical calculations~\ref{lQ} and~\ref{lU} show that the choice of a scanning strategy with a high degree of redundancies has a large impact on the reduction of systematic effects such as the IP leakage from the HWP. A scanning strategy similar to the one employed by \LB allows a reduction of the power spectrum amplitude by a factor of approximately 10 with respect to the \Planck\ scanning strategy. The overall systematic effect is expected to be much larger for ground experiments with a limited possibility of high degree of spin-2 scanning cross-linking.


\section{Combined Multi-Detector Effect}
\label{sec:multi-det}

Some degree of cancellations of the instrumental polarization is expected to happen when combining maps obtained with different detector observations with different azimuthal angles around the rotation axis of the HWP and at the same angular distance $\Theta$ from this axis. In order to verify this analytically, let's define $\Phi$, the azimuthal angle of the detector location in a fixed frame in the focal plane (see Figure~\ref{defInstr}). Let's come back to the definition of angles to see how the Mueller matrix coefficients are modified for any angle $\Phi$. The phases of the Mueller matrix coefficients $\phi_{ij}^{(nf)}$ (introduced in Equations~\ref{M2f} and~\ref{M4f} as $\phi_{ij}^{(2f)}$ and $\phi_{ij}^{(4f)}$) must be defined with respect to a reference axis in the focal plane coordinates. For the simulations used in this analysis, the phases $\phi_{ij}^{(nf)}$ are defined for a reference frame such that if $\rho=\psi$, i.e. the axis of the HWP and those of the focal plane are aligned, the incident wave vector is orthogonal to ordinary axis of the HWP. In other words the reference detector is located along one reference axis of the HWP: $\Phi=0$.

Now for a different location $\Phi\neq 0$, we can modify the coefficients of the Mueller matrix by rotating the frame of reference in the focal plane by $\Phi$: $\psi'=\psi+\Phi$ such that when $\rho=\psi'$ the detector observes with an incident angle orthogonal to the ordinary axis. The orientation of the polarizer in this new frame becomes $\psi_0'=\psi_0-\Phi$. All the previously derived expressions can be used by replacing $\psi$ by $\psi'$ and $\psi_0$ by $\psi_0'$. Consequently, considering the expressions of the leakage in~\ref{Prediction}, two detectors with azimuthal angles differing by 90 degrees would observe leakage $\delta Q$ and $\delta U$ maps with opposite signs.

It is important to notice that the previous operation is not equivalent to just adding $\Phi$ to the angles $\phi_{QI}$ and $\phi_{UI}$ in the expression of the Mueller matrix, which would then provide wrong predictions (such as a cancellation of the leakage with detectors 45 degrees apart as our intuition, since it is a 4$f_{\rm HWP}$ term, would falsely indicate).

\begin{figure}
  \centering
  \includegraphics[width=0.48\columnwidth]{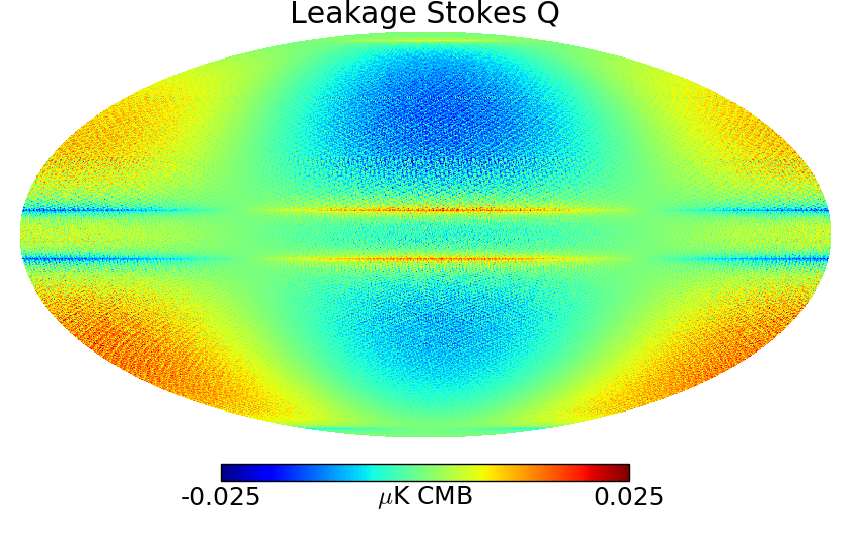}
  \includegraphics[width=0.48\columnwidth]{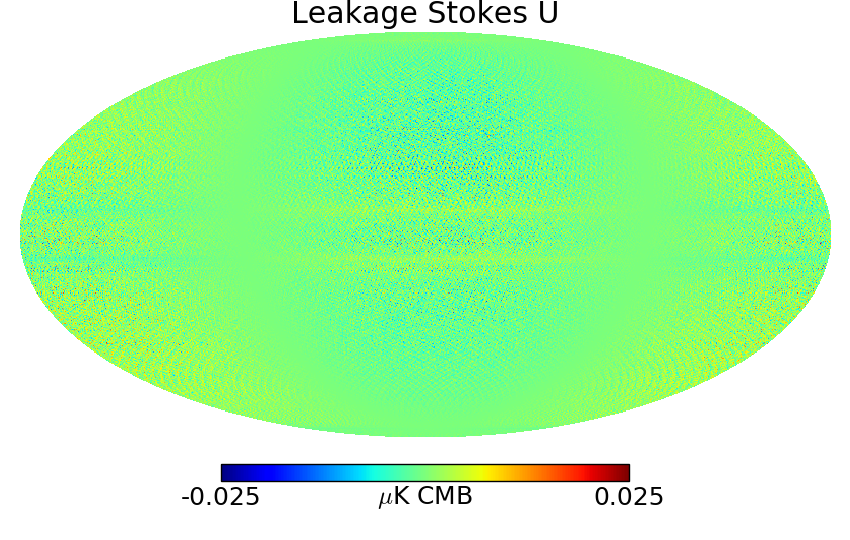}
  \includegraphics[width=0.48\columnwidth]{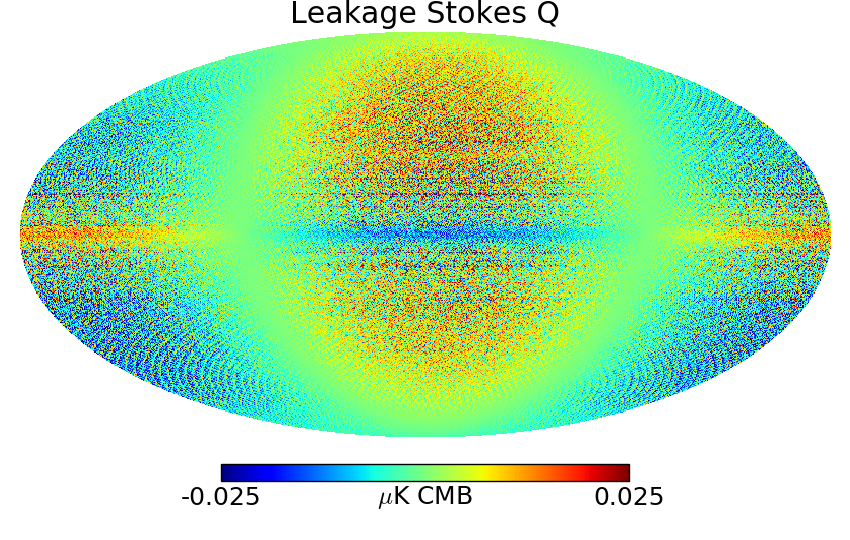}
  \includegraphics[width=0.48\columnwidth]{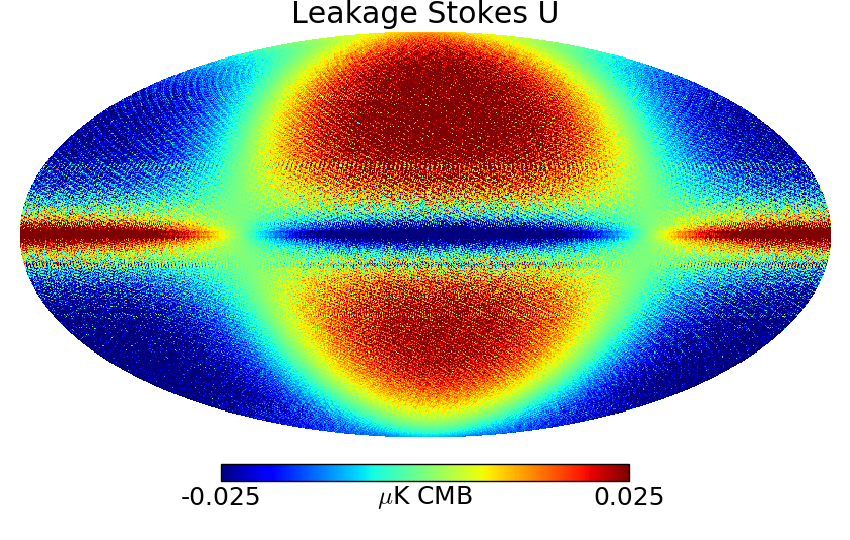}
  \caption{Leakage $\delta \widehat Q$ and $\delta \widehat U$ maps from the dipole only for two different detectors observing at different locations in the focal plane. Different phases are observed as well as different patterns related to the scanning strategy due to different rotation amplitude angles for the two detectors.}
    \label{SpSigboostleak}
\end{figure}

The detector to detector cancellations are not expected to be total since detectors observe at different locations in the focal plane, and the cross-linking parameters maps are significantly different for different corners of the focal plane~\footnote{Different locations along one of the axis in the focal plane are equivalent to different scanning strategy angle $\beta$, the observation angle with respect  to the rotation axis.}. Nevertheless cancellation should be more efficient at the largest scales. Figure~\ref{SpSigboostleak} show maps of leakage of the dipole including only the 4$f_{\rm HWP}$ component ($M^{(4f)}$ term in the Mueller matrix and not the $M^{(2f)}$ term) for detectors at different locations in the focal plane.
Different patterns for different detectors are observed depending on the phases of the projected IP as discussed previously, and on the differences in their scanning paths on the sky. The amplitude of the leakage depend on the incident angles $\Theta$ of the detectors with respect to the optical axis of the HWP. To calculate this we use the relation $\epsilon_1(\Theta)$ given in Equation~\ref{eq:modftheta}.

\begin{figure}
  \centering
  \includegraphics[width=0.7\columnwidth]{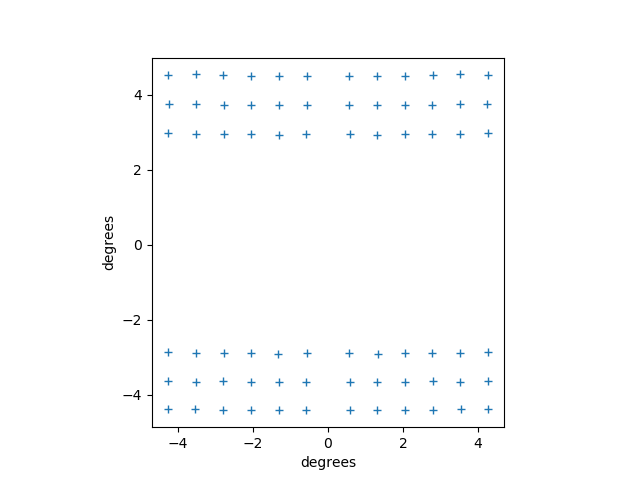}
  \caption{Detector locations in the LFT focal plane of the 140-GHz band assumed in this paper. This matches the definition of~\cite{Sekimoto21}}
    \label{FP}
\end{figure}

\begin{figure}
  \centering
  \includegraphics[width=0.49\columnwidth]{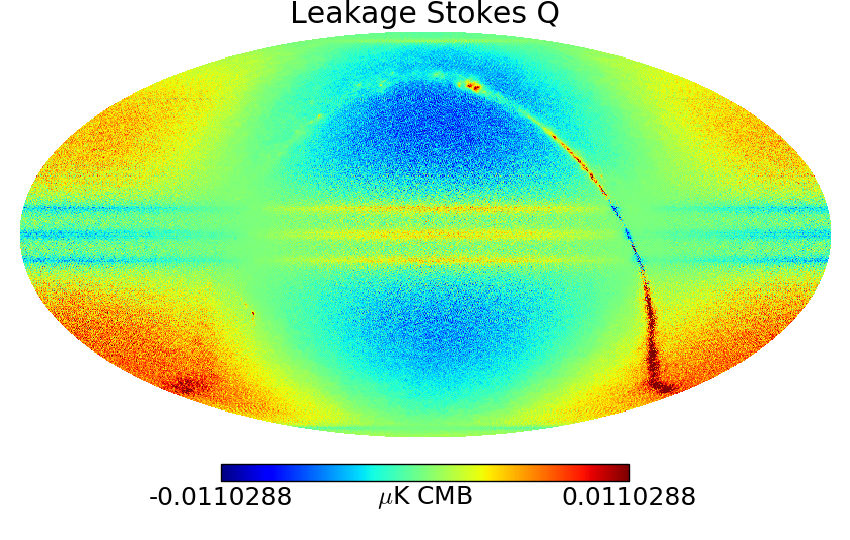}
  \includegraphics[width=0.49\columnwidth]{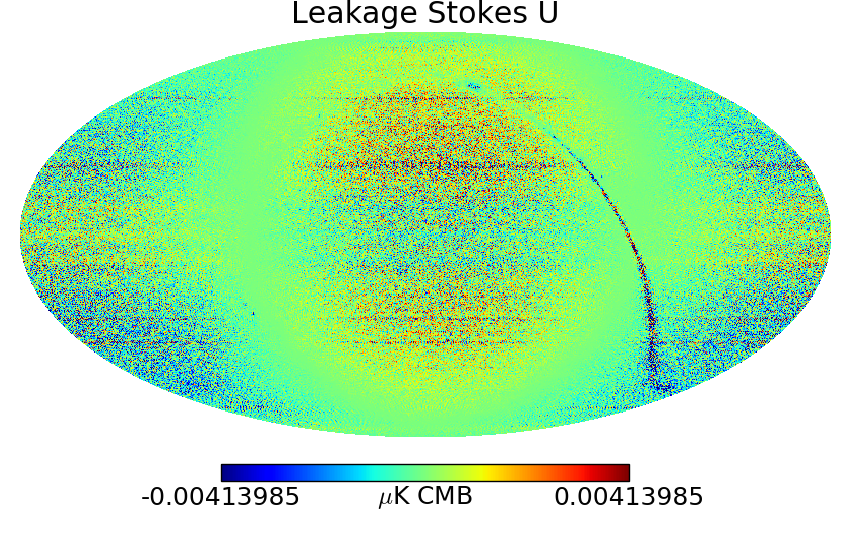}
  \includegraphics[width=0.49\columnwidth]{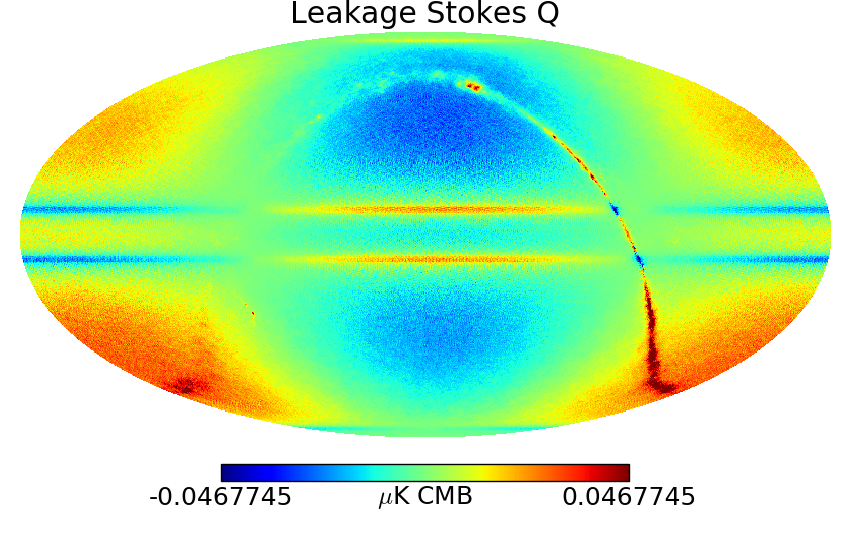}
  \includegraphics[width=0.49\columnwidth]{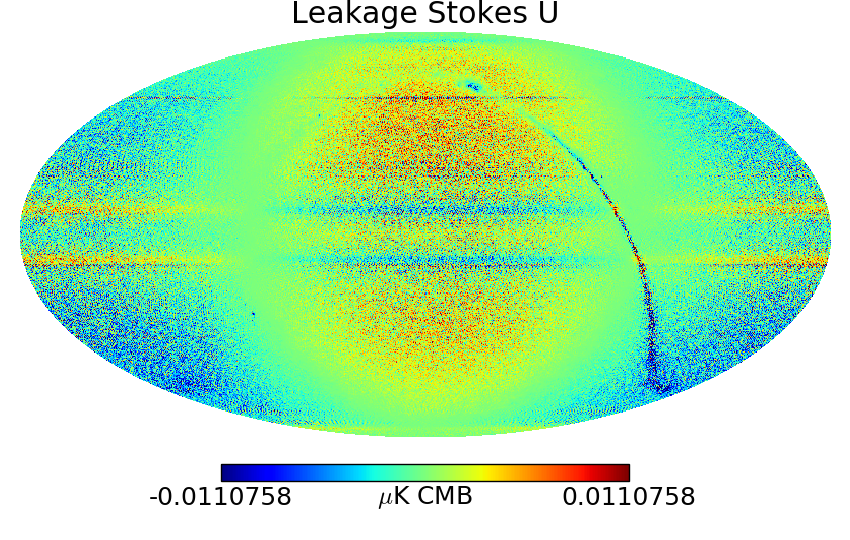}
  \caption{Stokes parameter $\delta \widehat Q$ and $\delta \widehat U$ maps of the $4f_{\rm HWP}$ IP leakage after combining maps obtained for all detectors of the 140-GHz band for the LFT focal plane configuration used in the design of \LB (as shown in Figure~\ref{FP}). Two optical configurations are studied: 1. the rotation axis of the HWP passes through the center of the focal plane (top row); 2. the HWP is tilted by 5 degrees (bottom row). The actual amplitude of the effect is positive for the $Q$ map and almost null for the $U$ map for the first configuration by chance since it depends on the phase of the IP effect and on the configuration of the detector locations in the focal plane. The amplitude of the total IP leakage is increased in case of a tilted HWP (notice the color scale for the two cases) because of less efficient canceling of the effect from detector to detector.}
    \label{combStokesIP}
\end{figure}

We have performed multi-detector timestream simulations at a single frequency band centered at 140\,GHz using a focal plane configuration of the low-frequency telescope of the \LB mission as indicated in Figure~\ref{FP} with different configurations of HWP optical axis position in the focal plane as we will specify later. A total of 144 detectors is assumed, two detectors with orthogonal polarizer angles are placed in the same physical location. Polarizers are oriented such that they are successively turned by 45 degrees from one adjacent detector pair to another. The polarizer orientation distribution in the focal plane has very little impact on the final results because the IP residual for each individual detector is independent of the angle $\psi_0$, and the optimal weights on each detector for the estimation of the final polarization maps are almost independent of this distribution.
Indeed, the final Stokes parameter maps are obtained by combining Stokes parameter maps from individual detectors following:
\begin{align}
{\bar Q}(p) = \frac{\sum n_i(p) {\widehat Q}_i(p)}{\sum n_i(p)}\label{mQ}\\
{\bar U}(p) = \frac{\sum n_i(p) {\widehat U}_i(p)}{\sum n_i(p)}\label{mU},
\end{align}
with $n_i(p)$ the number of hits per pixel $p$ for the detector $i$ and ${\widehat Q}_i$ and ${\widehat U}_i$ are the Stokes parameter maps obtained for detector $i$ obtained with SANEPIC. This is a nearly optimal solution in case of uncorrelated noise between detectors, and, also because of the use of an HWP (which we assume ideal for the solution) since with a quasi-instantaneous estimation of Stokes parameters: ${\rm Var}\{{Q_i}(p)\} = {\rm Var}\{{U_i}(p)\} \propto (n_i(p))^{-1}$ (this can easily be derived from Equation~\ref{ProjEq}),
and along with a high degree of cross-linking of the scanning strategy, the cross-terms in the covariance between Stokes parameters at any location on the sky.
We have calculated the combined effect from HWP imperfections coming from the dipole and sky emission anisotropies after averaging all the $\widehat Q$ and $\widehat U$ Stokes parameter maps as in Equations~\ref{mQ} and~\ref{mU} for all envisaged detectors, and after subtracting the input polarization maps (multiplied by a coefficients which is adjusted to remove the effect of non-ideal polarization efficiency). The residual maps are shown in Figure~\ref{combStokesIP}. We have assumed two cases: 1. an optical configuration such that the rotation axis of the HWP passes through the center of the focal plane; 2. an optical configuration with a tilted HWP by $5^\circ$ along the vertical direction in Figure~\ref{FP} and in the orthogonal direction with respect to the scanning direction. We have also isolated the dominant effect of the IP by including $\epsilon_1$ and $\epsilon_2$ only in the $M^{(4f)}$ matrix without other imperfections which is equivalent to setting the input polarization to zero from the timestream simulations.

\begin{figure}
  \centering
  \includegraphics[width=0.495\columnwidth]{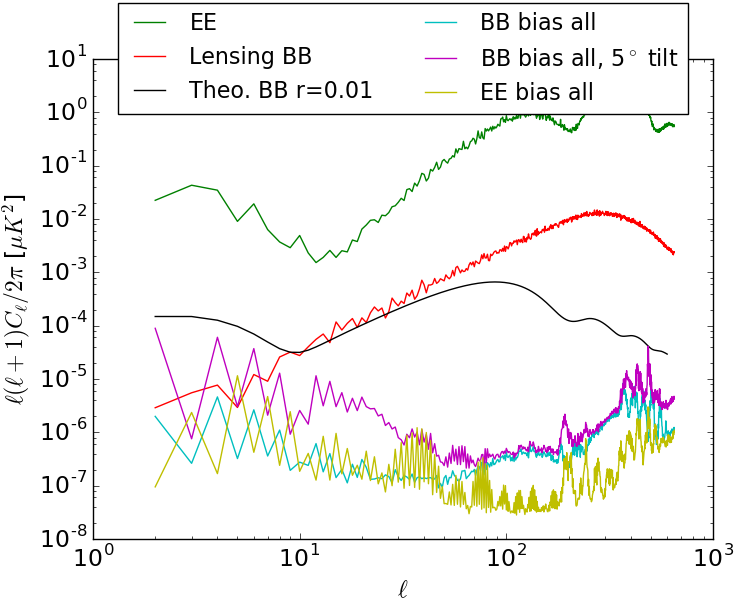}
  \includegraphics[width=0.495\columnwidth]{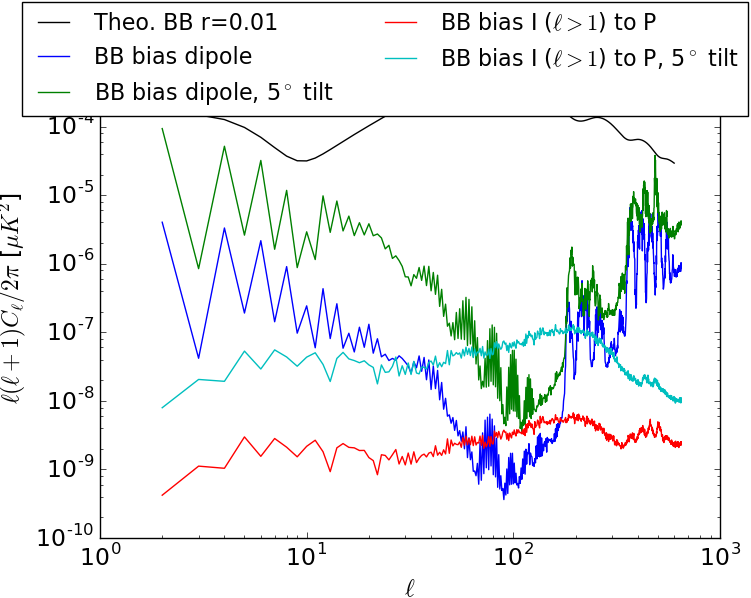}
  \caption{Left: Polarization $E$- and $B$-mode power spectra of the leakage maps $\delta \widehat Q$ and $\delta \widehat U$ for a model which includes all imperfections in the model of the Mueller matrix and combining all the detectors at 140-GHz LFT band envisaged for \LBm. The polarization $BB$ power spectra are shown for the two configurations of the HWP rotation axis. Right: Contribution from the IP of the dipole and CMB + foreground anisotropies.}
    \label{SpCombleak}
\end{figure}

We have estimated the combined effect on the polarization power spectra assuming the two focal configurations (with and without tilt). Those are shown in Figure~\ref{SpCombleak}. The IP leakage in the combined maps is greatly reduced as compared to contributions for individual detectors (see Figure~\ref{SpSigleak}). We can see that in case of a tilted HWP, the residual IP is boosted. This is expected because the detector to detector canceling of the IP discussed above is then less efficient due to the broken symmetry, and also, on average the detectors are located at larger incident angles.

The effect of $Q/U$ mixing induced by non-ideal corresponding coefficients of the Mueller matrix does not average when combining the Stokes parameter maps from different detectors, and is also present near normal incidence on the HWP. A large fraction of this effect can be accounted for by an angle shift $\rho+\delta\rho$ with respect to the assumed position of the HWP~\footnote{A related study is addressed in~\cite{Giardiello21} and requirements on this parameters can be found in~\cite{Krachmalnicoff21}}. This effective angle shift is related to the phases $\phi^{(4f)}_{XY}$ with $X$ and $Y$ being $Q$ and/or $U$, and is, in our model, the same for all the detectors independently of their location in the focal plane. We can clearly see the effect in the $EE$ and $BB$ power spectra of the residual maps $\delta Q$ and $\delta U$, which are the result of the combined contributions from all the detectors of systematic effect, at multipoles in the range $60 \lesssim \ell \lesssim 500$. This effect appears with the same amplitude with or without the tilt of the HWP, while it was not visible in power spectra for individual detectors due to higher IP leakage. We observe the leakage from $E$ to $B$ modes of the CMB with patterns in the $B$-mode power spectrum mimicking the oscillations of the $E$ mode power spectrum. The Galactic foregrounds have negligible contributions, thanks to the Galactic mask selecting 50\,\% of the sky for the analysis. 

It is clear that the position of detectors observing at the same frequency in the focal plane can then be optimized to minimize the total effect of the IP by the HWP on the final polarization Stokes parameters. This is an important fact to consider for the optimization of a mission using a HWP.

The analysis of the impact of IP accounting for the observations at all the frequency channels, involving the propagation of the effect through component separation, as well as the modeling of frequency dependence of the HWP Mueller matrix coefficients, is postponed to a future study.

We have computed the impact of the IP to the measurement of the tensor-to-scalar ratio $r$ by fitting the amplitude $\delta r$ of the model of the $B$-mode power spectrum to the residual power spectrum. The value $\delta r$ represents the bias on $r$ induced by the leakage and is obtained by maximizing the log-likelihood of the data: 
\begin{equation}
\log L(r) = - f_{sky}\frac{1}{2}\sum_{\ell=2}^{\ell_{\rm max}}(2\ell+1)\left[\frac{{\widehat C}_\ell}{C_\ell}+\log C_\ell -\frac{2\ell-1}{2\ell+1}\log {\widehat C}_\ell-1\right].
\end{equation}
In this expression $C_\ell$ the theoretical $B$-mode power spectrum modeled as:
\begin{equation}
C_\ell = r C_\ell^{\rm tens} + C_\ell^{\rm lens} + N_\ell + C_\ell^{\rm foreg},
\end{equation}
with $C_\ell^{\rm tens}$ the primordial $B$-mode spectrum for $r=1$; $C_\ell^{\rm lens}$ the lensing power spectrum; $N_\ell$ the expected level of noise $B$-mode power spectrum for \LB observed maps; $C_\ell^{\rm foreg}$ the expected residual foreground power spectrum after an subtraction of the bias induced by foreground separation~\cite{Errard19}; and ${\widehat C}_\ell$ the model of the observed spectrum including the systematics:
\begin{equation}
{\widehat C_\ell} = C_\ell^{\rm sys} + C_\ell^{\rm lens} + N_\ell + C_\ell^{\rm foreg},
\label{Eq:Cells}
\end{equation}
with $C_\ell^{\rm sys}$ the power spectrum of HWP systematics.
We have assumed no primordial spectrum ($r=0$) so we obtain $\delta r$ by setting:
\begin{equation}
\frac{\partial L(r)}{\partial r}_{|_{r=\delta r}} = 0
\label{Eq:dLdr}
\end{equation}

\begin{table}[htbp!]
 \centering
			\begin{tabular}{|c|c|c|c|c|}
				\hline
				& HWP centered & HWP tilted by 5$^\circ$  & Mitigated, model~1 & Mitigated, model~2\\
				\hline
				\hline
				$\bar{\epsilon_1}$ & 2.23 $\times 10^{-5}$ & 4.38 $\times 10^{-5}$ &  &   \\
				\hline
				$\delta r$ & $1.36 \times 10^{-4}$ & $1.47 \times 10^{-3}$ &  $< 1.20 \times 10^{-6}$ & $< 1.84 \times 10^{-7}$ \\
				\hline
			\end{tabular}
\caption{Contribution of the IP to the tensor-to-scalar ratio bias $\delta r$. The values of $\delta r$ quoted for the first two columns correspond to the bias induced by the HWP imperfections for the whole 140-GHz band of the LFT instrument. The other clumns are for after correction for 3 year observations. The most pessimistic case of a tilted HWP is used for evaluation of the effect after correction with two different schemes: for model~1, when the $\epsilon_i$ coefficients are estimated independently for each detector; and for model~2 when we use the scaling relation $\epsilon_1(\Theta, \phi)$ accounting for the exact relative phase dependence of this parameter with respect to the azimuthal angle of the detector locations $\phi$ around the HWP rotation axis. We also indicate the mean IP parameter for the full frequency channel.
\label{tab:drIP}
}
\end{table}

The evaluated bias for the different configuration of the HWP (tilt by 5$^\circ$ or no tilt) are shown in Table~\ref{tab:drIP}. We have assumed the foreground residual power spectra $C_\ell^{\rm foreg}$ shown in~\cite{PTEP} from the method developed in~\cite{Stompor09}. The noise power spectrum $N_\ell$ is fixed to the expected spectrum for three year mission length combining all detectors, although we use the residual IP power spectrum of the 140-GHz LFT band only. This is justified by the fact that we are aiming at evaluating the contribution of the 140-GHz band to the final likelihood.
Using the 140-GHz band noise spectrum increases by a small amount the bias because it gives more weight to the reionisation peak in the likelihood where the foreground residuals dominate in the total uncertainty.

\begin{figure}
  \centering
  \includegraphics[width=0.6\columnwidth]{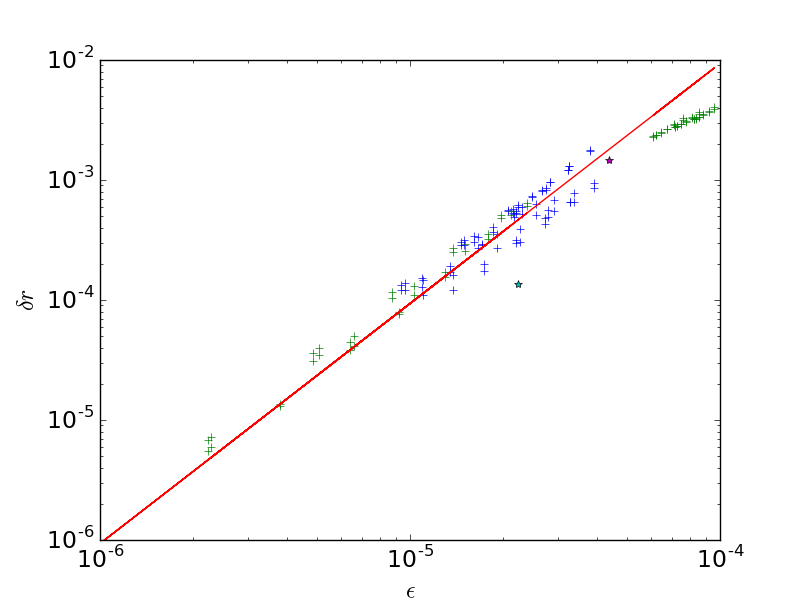}
  \caption{Fitted tensor-to-scalar ratio $\delta r_i$ from the residual maps for each individual detector versus the IP parameter $\epsilon_{1;i}$. The blue crosses are for no HWP tilt and the green crosses for the same focal plane but with a tilted HWP by 5 degrees. The values for the full focal plane are indicated as stars. The red line shows a quadratic relation.}
    \label{Fig:Alldr}
\end{figure}

In Figure~\ref{Fig:Alldr} we show the calculated the fitted tensor-to-scalar ratio obtained with Equation~\ref{Eq:dLdr} from individual detector maps by inserting in Equation~\ref{Eq:Cells} as $C_\ell^{\rm sys}$ the power spectrum of the individual contributions to the residual IP. We observe deviations from the quadratic relation between the IP parameter $\epsilon$ and $\delta r$ which are attributed to the differences of phases of the IP, reshuffling the power between $E$, $B$ modes and the different multipoles (the phase depends on the location in the focal plane). We also indicate values of $\delta r$ for the full frequency band and the two configurations, at the location of the mean coefficients in the focal plane. The can see that the cancellation of the effect by combining detector maps is more efficient for the non-tilted HWP configuration.

The multi-frequency study of the effect of instrumental polarization by the HWP is postponed to a future publication. This analysis for the \LB experiment would require: 1. coherent simulations of the coefficients of the three HWP across all the HWP frequency bands (between about 30 to 150\,GHz for LFT); 2. coherent multi-frequency simulations; 3. propagation of the residual systematic effect through component separation after map-making. Some expected results will be discussed at the end of the next part, but we can already state that we expect a lower level of IP for other LFT CMB frequencies since the 140-GHz band is at the edge of the HWP functioning band.

\section{Correction of the Instrumental Polarization}
\label{sec:correction}

We propose a method to estimate the instrumental polarization parameters for a given frequency band from the science data and to correct the data exploiting the predictable effect on the large dipole signal (and possibly the Galactic signal) jointly with the Stokes parameter maps.

We develop a formalism to fit the parameters related to the $4f_{\rm HWP}$ component of the Mueller matrix: $\epsilon_1(\Theta)$, and $\phi_{QI}$ using the physical relations derived from previous studies: $\phi_{QI} = \phi_{UI} + \frac{\pi}{2}$ and $\epsilon_1 = \epsilon_2$. The parameters related to the other components, $0f_{\rm HWP}$ and $2f_{\rm HWP}$ are not accounted for since the projected leakage on the final maps is small (or does not induce a mismatch) although we include these parameters in the simulations. The parameter $\epsilon_1$ is estimated at an arbitrary reference angle $\Theta_{\rm ref}$ and the model derived in Equations~\ref{eq:epsThetaMod} and~\ref{eq:modftheta} is used to express $\epsilon_1$ with respect to any incident angle $\Theta$. So within this framework, $a$, $\epsilon_1(\Theta_{\rm ref})$ and $\phi_{QI}$ constitute the set of parameters to estimate as well as the Stokes parameter maps.

Notice that all the relevant quantities used in this section are recalled in Table~\ref{tab:variables}.  

\subsection{Model and Maximum Likelihood Solution}

The model of the data could be written based on the previous section, but in a more convenient way:
\begin{equation}
d_{it} = {\cal M}_{itpc}\, m_{pc} + \alpha_i o_t + n_{it}.
\label{eq:modelCorr}
\end{equation}
We use by convention in this paper, implicit summation for repeated indices. In the previous relation $d_{it}$ is one element of the time-stream data vector for detector $i$; $n_{it}$ is one element of the noise vector; $m_p = \{I_p,Q_p,U_p\}$ is a vector containing the Stokes parameters (numbered by $c$) in the sky coordinates for every pixel in the map; $o$ is the orbital dipole due to the motion of the Satellite around the Sun; $\alpha_i$ is a calibration factor which depends on elements of the Mueller matrix\footnote{The orbital dipole is added here for a complete model and is useful to consider as it is a calibrator. For the following, since we do not fit for the calibration coefficient, we can assume that we can remove it from the data and we do not consider it in this paper.}; and finally, ${\cal M}_{pit}$ is a vector which incorporates the Mueller matrix and the pointing matrix such that, based on Equation~\ref{eq:modelobs}, for a given element corresponding to pixel $p$ observed at time $t$:
\begin{equation}
 {\cal M}_{itp} = \Big(1 \,\, \cos(2{\psi_0}_i-2\Phi_i) \,\, \sin(2{\psi_0}_i-2\Phi_i)\Big)\, M(\Theta_i,\rho_t-\psi_t-\Phi_i)\, R(2\psi_t+2\Phi_i)
 \label{eq:CalM}
\end{equation}
where we have accounted for different detector locations as discussed in Section~\ref{sec:multi-det} with the azimuthal angle $\Phi_i$ of detector $i$ in focal plane coordinates, and $R$ is the rotation matrix, $M$ the HWP Mueller matrix in the focal plane coordinates. The index $c$ is not indicated in the expression above since ${\cal M}_{itp}$ is a three-component vector encoding data for the three Stokes vectors. The matrix ${\cal M}$ is very sparse as usual since only elements corresponding to pixel $p$ observed at time $t$ are different than 0.
In the simplified case for which only $4f_{\rm HWP}$ IP parameters are assumed, and neglecting the impact of all other HWP imperfections, the model~\ref{eq:modelCorr} gives for the contribution of imperfection to the data:
\begin{equation}
\delta d_{it} = \epsilon_i \cos(4\rho_t - 4\psi_t - 2\Phi_i - 2{\psi_0}_i + \phi_{QI})\, I_{p(t)},
\end{equation}
after accounting for a general detector location in the focal plane, assuming the amplitude of the 4$f$ components $\epsilon_{1;i}=\epsilon_{2;i}=\epsilon_i$, and keeping the relation between the phases $\phi_{QI}$ and $\phi_{UI}$ as previously discussed.

Changing the parametrization we can write:
\begin{equation}
\delta d_{it} = [\lambda^{(1)}_i \cos\zeta_{it} + \lambda^{(2)}_i \sin\zeta_{it}] \,(I_{p(t)} - {\bar{I}}) + [\lambda^{(3)}_i \cos\zeta_{it} + \lambda^{(4)}_i \sin\zeta_{it}].
\end{equation}
with
\begin{equation}
\zeta_{it} \equiv 4\rho_t - 4\psi_t - 2\Phi_i - 2{\psi_0}_i,
\end{equation}
with the two amplitudes $\{\lambda_i^{(1)};\lambda_i^{(2)}\}$ directly related to $\epsilon_i$ and the phase $\phi_{QI}$ by
\begin{align}
\lambda_i^{(1)}&=\epsilon_i \cos(\phi_{QI})\\
\lambda_i^{(2)}&=-\epsilon_i \sin(\phi_{QI}).
\label{eq:modlamb}
\end{align}
We have subtracted the mean intensity ${\bar{I}}$ from $I_{p(t)}$ in the expression above because the mean intensity is inaccessible from the data, as opposed to the dipole when measuring the Stokes parameter $I$ map. However, the effect of $I$ on the polarization map due to the IP leakage depends on the monopole. We have added two extra free parameters $\lambda_i^{(3)}={\bar{I}}\,\lambda_i^{(1)};\, \lambda_i^{(4)}={\bar{I}}\,\lambda_i^{(2)}$ to absorb mismatches in the intensity map estimation due to the monopole.

This reparametrization allows to linearize the equations with respect to the unknown parameters since $\zeta_{it}$ is a known quantity and depends only on the scanning strategy while $\phi_{QI}$ have realistically to be assessed from the data and is obtained from the estimation of $\lambda_i^{(1)}$ and $\lambda_i^{(1)}$ with our approach as detailed later. Consequently we can define a set of four templates to fit in the timestreams:
\begin{align}
	T_{itq=1} &= (I_{p(t)} - {\bar{I}})\cos\zeta_{it}\, ;\,\, T_{itq=2} = (I_{p(t)} - {\bar{I}})\sin\zeta_{it} ;\\ T_{itq=3} &= \cos\zeta_{it}\, ;\,\, T_{itq=4} = \sin\zeta_{it},
	\label{Eq:Ttemplate}
\end{align}
and the amplitudes $\lambda_i = \{\lambda_i^{(1)};\lambda_i^{(2)};\lambda_i^{(3)};\lambda_i^{(4)}\}$ would be the unknown parameters. Let's notice that this requires an estimation of the intensity $I_{p(t)}$. The cosmological dipole as measured by \Planck\ satellite~\cite{PlanckResults18} could be used as a template, but more accurately, the total intensity can be estimated iteratively from the recovered Stokes parameters as we will see below. Let's notice that in case the intensity $I$ is independent of time, as for the CMB monopole, only $\lambda_i^{(3)}$ and $\lambda_i^{(4)}$ are different than 0 and we can anticipate that the optimal solution for those parameters will be simply obtained by extracting the Fourier coefficients of the residual data at $4f_{\rm HWP}$ since the corresponding functions $T$ would be Fourier basis functions.

For the following, let's rewrite the data model including the new parametrization, removing as well the orbital dipole contribution:
\begin{equation}
d_{it} = {\bar{\cal M}}_{itpc} \,m_{pc} + T_{itq}\lambda_i^{(q)}  + n_{it}
\label{eq:modelCorr2}
\end{equation}
where ${\bar{\cal M}}$ is similar to ${\cal M}$ but with the IP terms in the Mueller matrix set to 0; and $c$ takes values in the interval $\{0,1,2\}$. Notice that in the previous equation the index $i$ is repeated in the second term of the right hand side but not summed over. In case no external templates are used, the term $T$ contains a dependence on the Stokes parameters $m$ since $I_t$ in Eq.~\ref{Eq:Ttemplate} is built from the recovered intensity map as
\begin{equation}
    I_{p(t)} = {\bar{\cal M}}_{itpc=0} \,m_{pc=0}
    \label{Ipt}
\end{equation}

Based on the data model, we can write a criterion derived from the log likelihood of the data assuming Gaussian stationary noise in timestream data, and independence of noise between different detectors
\begin{equation}
-2\log{\cal L} = \sum_i \Big[d_{it} - {\cal M}_{itpc} m_{pc} \Big]^tN_{itt'}^{-1}\Big[d_{it'} - {\cal M}_{it'pc} m_{pc}\Big] + K.
\end{equation}
For the following, we do not write the summation over pixels, time samples and Stokes parameter components which are hidden in matrix operations reducing the model expression to (vector quantities are indicated in bold font):
\begin{equation}
-2\log{\cal L} = \sum_i \Big[\vecd_i - {\cal M}_i \vecm \Big]^tN_i^{-1}\Big[\vecd_i - {\cal M}_i \vecm\Big] + K.
\end{equation}
In the previous expressions $K$ is a independent of the parameters and $N_i=<\vecn_i\vecn_i^t>$ is the noise covariance matrix for detector $i$. We assume that the $\alpha$ parameters are known for simplicity (those are related to the detector calibration and constrained by this procedure), but our approach can be generalized for the estimation of those parameters.
We maximize ${\cal L}$ with respect to $m$ and $\lambda_i$:
\begin{equation}
\frac{\partial {\cal L}}{\partial \vecm}_{|_{\widehat{\veclambda}}} = 0;\,\,\, \frac{\partial {\cal L}}{\partial \veclambda_i}_{|_{\widehat{\vecm}}} = 0
\end{equation}
This gives the coupled set of equations (removing indices which are summed over):
\begin{eqnarray}
  \widehat{\vecm} = \left[\sum_i{\cal M}_i({\widehat{\veclambda_i}})^tN_i^{-1}{\cal M}_i({\widehat{\veclambda_i}})\right]^{-1} \sum_i{\cal M}_i({\widehat{\veclambda_i}})^tN_i^{-1} \vecd_i\label{eq:mmStockes_}\\
  \widehat{\veclambda}_i = [{\widehat T}_i^tN_i^{-1}{\widehat T}_i]^{-1} {\widehat T}_i^tN_i^{-1} (\vecd_i -{\bar{\cal M}}_i\widehat{\vecm}) \label{eq:mmpHWParams},
\end{eqnarray}
The first relation is similar to the well known map-making equation (see e.g~\cite{Patanchon08,DeGasparis16}), but in our case it incorporates Mueller matrix coefficients coupling the Stokes parameter coefficients. We have indicated ${\widehat .}$ for the template $T_i$ because it is built using the estimated intensity as stated before.
This formalism does not make any assumption about the model relating the different $\lambda_i$ for all detectors, but we know they are linked by the model of the HWP (since the IP is a property of the HWP common to all detectors), in particular by the phase relation~\ref{eq:modlamb} ($\phi_{QI}$ is common to all detectors) and if available by a model of the dependence of $\epsilon$ with $\Theta$ as in Equation~\ref{eq:modftheta}. Ideally, we would estimate directly, instead of all the parameters $\widehat{\lambda}_i$, as in Equation~\ref{eq:mmpHWParams}, the parameters $a$, $\epsilon(\Theta_{\rm ref})$ and $\phi_{QI}$. In this case, direct minimization would not lead to linear equations and would need to be performed iteratively. In an approach where the individual ${\lambda}_i$ are estimated without using the model, iterating between relations~\ref{eq:mmStockes_} and \ref{eq:mmpHWParams}, we can expect a degradation of the performance in the correction of the IP parameters as compared to an approach which would use the model. The method above with free $\lambda_i$ is somewhat conservative. We introduce such an approach with model constraint in the following.

First of all, it is clear that the method can not provide tight constraints on the parameter $a$ since the ratio of the parametric function (describing the IP dependence with respect to the incident angle) evaluated at two angles varies only weakly with respect to $a$ for small incident angles. Then in practice we can fix the parameter $a$ to the best guess from RCWA simulations. We define new templates for each detector linked together with the rescaling function $f$ as the following:
\begin{align}
 Z_{itq=1} &= \frac{f(\Theta)}{f(\Theta_{\rm ref})} \cos\zeta_{it}\,(I_{p(t)} - {\bar{I}}); \,\,\, Z_{itq=2} = \frac{f(\Theta)}{f(\Theta_{\rm ref})} \sin\zeta_{it}\,(I_{p(t)} - {\bar{I}}); \\ Z_{itq=3}&= \frac{f(\Theta)}{f(\Theta_{\rm ref})} \cos\zeta_{it}; \,\,\, Z_{itq=4} = \frac{f(\Theta)}{f(\Theta_{\rm ref})} \sin\zeta_{it}.
\end{align}
Now the model of the observation using the model constraint writes:
\begin{equation}
d_{it} = {\bar{\cal M}}_{itpc} \,m_{pc} + Z_{itq} \lambda^{(q)}  + n_{it},
\label{eq:modelCorr3}
\end{equation}
where the parameters $\lambda^{(q)}$ for $q=\{1,2,3,4\}$, common to all detectors, are equal to $\lambda^{(q)}_i$ for an arbitrary detector at the reference incident angle $\Theta_{\rm ref}$ and at a reference phase $\Phi_{\rm ref}=0$.

The maximum likelihood solution for $\lambda$ as well as the maps would then be (again removing indices corresponding to Stokes parameter components, pixels, and time stamps which are summed over by matrix operations):
\begin{eqnarray}
\widehat{\vecm} = \left[\sum_i{\cal M}_i({\widehat{\veclambda}})^tN_i^{-1}{\cal M}_i({\widehat{\veclambda}})\right]^{-1} \sum_i{\cal M}_i({\widehat{\veclambda}})^tN_i^{-1} \vecd_i\label{eq:mmStockes}\\
\widehat{\veclambda} = \left[\sum_i {\widehat Z}_i^tN_i^{-1}{\widehat Z}_i\right]^{-1} \sum_i {\widehat Z}_i^tN_i^{-1} (\vecd_i -{\bar{\cal M}}_i\widehat{\vecm})\label{eq:mmpHWParAve}
\end{eqnarray}
where, for the first expression, ${\cal M}_i({\widehat{\veclambda}})$ is derived from the model~\ref{eq:CalM} using the set of fitted HWP parameters, i.e. the derived $\epsilon_i$ and $\phi_{QI}$ from $\widehat{\veclambda}$. The parameters of our HWP model are then simply\footnote{Note that $\lambda^{(3)}$ and $\lambda^{(4)}$ also carry information on $\phi_{QI}$ but with significantly less precision since the uncertainties on those parameters are dominated by those on $\bar{I}$.}:
\begin{equation}
\epsilon(\Theta_{\rm ref}) = \sqrt{(\lambda^{(1)})^2 + (\lambda^{(2)})^2};\,\,\,\phi_{QI} = \arctan\left(-\frac{\lambda^{(2)}}{\lambda^{(1)}}\right).
\end{equation}

The form~\ref{eq:mmpHWParams}, and the approach of independent $\lambda_i$ estimation described in the first place, are useful in case a more general model is used for the dependence of IP with the incident angle.

Let's make two remarks: 

1. For the resolution of the map-making equation~\ref{eq:mmStockes}, one could use a simple projection algorithm (i.e $N_i$ diagonal). This would be nearly optimal with a perfect HWP for the recovery of polarization Stokes parameters, but we expect to degrade the performance of the method since: the $1/f$ noise is expected to slightly leak into the polarization Stokes parameters, precisely due to the IP; and the recovery of the intensity used for the template $T_i$ reconstruction would be degraded.

2. Regarding the second relation~\ref{eq:mmpHWParAve}, in case of constant $I$ and assuming a constant rotation velocity of the HWP, the solution is identical to the Fourier decomposition coefficient of the residual data at $4f_{\rm HWP}$, after removing the estimated map, since the functions $T$ are Fourier modes provided that 
the noise matrix $N$ is diagonal in Fourier space. It is then independent of the noise matrix in case of stationary noise:
\begin{equation}
  \widehat{\lambda}^{(1)}_i + \sqrt{-1}\, \widehat{\lambda}^{(2)}_i\propto {\cal F}\{\vecd_i -{\bar{\cal M}}_i\widehat{\vecm}\}(4f_{\rm HWP})
\end{equation}
The solution can then obtained by simply selecting the mode at $4f_{\rm HWP}$ after Fourier transform, as we expected.

In the case studied in this paper using the dipole and CMB anisotropies, the resolution of~\ref{eq:mmpHWParams} is done quickly since it involves only the computation of the Fourier transform of the data once. The resolution then corresponds, in term of computation time, to one iteration of the GLS map-maker. The involved matrices in Fourier space are very sparse for slowing varying $I_{p(t)}$ as the dipole is the dominant signal.


\subsection{Approximate Maximum Likelihood Solution}

The resolution of~\ref{eq:mmStockes} requires the same resources than the GLS map-making. To solve this equation, the usual codes can be modified to incorporate the Mueller matrix ${\cal M}_i$ in place of the usual pointing matrix. However we adopt in this paper a simpler solution departing from the maximum likelihood solution to solve for the maps and avoiding modifications of the methods. We consider a two step approximation of the maximum likelihood for the solution. First we use:
\begin{equation}
    \widehat{\vecm}_{1} = \left[\sum_i[{\cal M}^{\rm Eff}_i]^tN_i^{-1}{\cal M}^{\rm Eff}_i\right]^{-1} \sum_i[{\cal M}^{\rm Eff}_i]^tN_i^{-1} (\vecd_i-\widehat{\vecP}_i)
    \label{eq:hm1}
\end{equation}
where we assume an effective HWP matrix for the solution ${\cal M}^{\rm Eff}$ built from Equation~\ref{eq:CalM} using the Mueller matrix $M^{\rm Ideal}$ defined in Equation~\ref{Mideal}, and accounting for an exact (i.e. perfectly known) polarization efficiency due to other Mueller matrix imperfections (however we do not account for the coupling between Stokes $Q$ and $U$ parameters). This matrix is used instead of $\bar{\cal M}$ as we will justify later. We have also defined $\widehat{P}_{it}=T_{itq}\widehat\lambda_i^{(q)}$ or $\widehat{P}_{it}=Z_{itq}\widehat \lambda^{(q)}$ depending on which set of parameters is estimated. Second we use:
\begin{equation}
    \widehat{\vecm}_{2} = \sum_i W_i \Bigg( \left[[{\cal M}^{\rm Eff}_i]^tN_i^{-1}{\cal M}^{\rm Eff}_i\right]^{-1} [{\cal M}^{\rm Eff}_i]^tN_i^{-1} (\vecd_i-\widehat{\vecP}_i) - {\overline{\widehat I_i}}\,\vecu \otimes \begin{pmatrix}
1 \\ 0 \\ 0
\end{pmatrix}\Bigg),
     \label{eq:hm2}
\end{equation}
with
\begin{equation}
    W_i  = {\rm diag} \left[[{\cal M}^{\rm Eff}_i]^tN_i^{-1}{\cal M}^{\rm Eff}_i\right] \Big[\sum_j {\rm diag} \left[[{\cal M}^{\rm Eff}_j]^tN_j^{-1}{\cal M}^{\rm Eff}_j\right]\Big]^{-1}.
     \label{eq:whm2},
\end{equation}
and $\vecu$ a single map size vector containing 1 as components for each pixel such that the operation $\vecu \otimes (1\,\,\, 0\,\,\, 0)^t$ builds a quantity in the extended pixel times 3 Stokes component space which has the value 1 for the intensity map component only. The quantity $[{\cal M}^{\rm Eff}_i]^tN_i^{-1}{\cal M}^{\rm Eff}_i$ is the inverse noise covariance matrix between the 3 Stokes parameter maps (of size $3N_p \times 3 N_p$), and the 'diag' operation extracts only the diagonal part of it.  
For the IP parameters we use:
\begin{equation}
\widehat{\veclambda}_i = [{\widehat T}_i^tN_i^{-1}{\widehat T}_i]^{-1} {\widehat T}_i^tN_i^{-1} (\vecd_i -{\cal M}^{\rm Eff}_i\widehat{\vecm}) \label{eq:mmpHWParams_ap},
\end{equation}
or
\begin{equation}
 \widehat{\veclambda} = \left[\sum_i {\widehat Z}_i^tN_i^{-1}{\widehat Z}_i\right]^{-1} \sum_i {\widehat Z}_i^tN_i^{-1} (\vecd_i -{\cal M}^{\rm Eff}_i\widehat{\vecm})\label{eq:mmpHWParams2_ap},
\end{equation}
depending on the model used.

In the first equation (\ref{eq:hm1}) we have used the effective HWP Mueller matrix ${\cal M}^{\rm Eff}$ instead of $\bar{\cal M}$, simulating the case we ignore other imperfections such as 2$f_{\rm HWP}$ or imperfections leading to Q/U mixing (while those are included in our simulations). Using the matrix $\bar{\cal M}$ would give the solution after maximization of the likelihood if the templates $T_{itq}$ and $Z_{itq}$ were independent of the map parameters, but we have seen they are build from the recovered intensity $\widehat{m}_0$. Because of this dependence this solution is not equivalent to Equations~\ref{eq:mmStockes_} (or \ref{eq:mmStockes}). In practice, we choose to calculate the templates from the recovered intensity maps at the preceding iteration. We do not expect a priori a significant impact of the approximate solution for the maximum likelihood on the final performance of the method because the intensity is estimated with very high signal to noise with \LBm, and errors made on the template have a very low impact on the IP parameter estimation noise and consequently on the recovered maps $\widehat m$ (this is confirmed by the very few iterations needed between the estimation of $m$ and $\lambda$ in order to reach the convergence).
The second solution $\widehat{m}_{2}$ is implemented for our analysis. As $W_i$ is a diagonal matrix, this solution is the weighted sum of stokes parameter maps obtained for individual detectors as for Equations~\ref{mQ} and~\ref{mU}, each of the individual maps (whose expression is within large parenthesis) being the maximum likelihood solution for individual detectors. We also subtract the mean of the individual recovered intensity map ${\overline{\widehat I_i}}$ from the recovered map of each detector since those are not constrained by individual detector data and could bias the combined maps due to the non-uniform weights $W_i$. This operation affects (and almost nulls) the monopole of the recovered intensity map but it does not bias the recovered IP parameters $\lambda_i^{(1)}$ and $\lambda_i^{(2)}$ since any change in the monopole is absorbed by $\lambda_i^{(3)}$ and $\lambda_i^{(4)}$.

Finally, we claim that $\widehat{m}_{2}$ is a relatively good approximation of the global maximum likelihood solution because: 1. different detectors have similar scanning strategies, so the extra redundancies brought by the joint analysis of detectors to remove 1/f noise from intensity maps is limited; 2. there is low coupling between Stokes $I, Q, U$ parameters thanks to the HWP 3. 1/f noise does not project in $Q$ and $U$ Stokes parameter maps leading to nearly diagonal associated covariance matrix ($[{\cal M}^{\rm Eff}_i]^tN_i^{-1}{\cal M}^{\rm Eff}_i$ is nearly diagonal except for the intensity sub-matrix part). This is supported by internal studies within the \LB collaboration and will be the object of a future publication.

The implementation of~\ref{eq:mmStockes_} and \ref{eq:mmStockes} requiring modifications of the map-making code, is postponed to a future study which would include as well other types of imperfections of the Mueller matrix such as coefficients producing $Q$ and $U$ mixing.

For the second step of each iteration, the solutions~\ref{eq:mmpHWParams_ap} or \ref{eq:mmpHWParams2_ap} are implemented, with ${\cal M}^{\rm Eff}$ for the IP-free Mueller matrix. For simplicity we have assumed a diagonal matrix for the noise matrix $N$. This has a negligible impact on the final results since the $T$ and $Z$ templates have a very narrow band spectrum centered at $4f_{\rm HWP}$, as the templates are made of a nearly single frequency sine wave modulated with a slowing varying function (the intensity is dominated by the dipole signal), and the noise covariance matrix is diagonal in Fourier space such that only the noise at $4f_{\rm HWP}$, located in the white part, matters. The intensity timestream used to build the templates are reconstructed applying the relation~\ref{Ipt} which is equivalent to reading the recovered intensity map using the detector pointing.

All the approximations discussed above are justified by the very low observed residuals after correction shown in the next section. 

\subsection{Performance of the Correction and Discussion}

We have simulated the observations with a non-ideal HWP incorporating all the modeled defects in the Mueller matrix, i.e. the $2f_{\rm HWP}$ and $4f_{\rm HWP}$ terms as well as the perturbations for all the coefficients (including the $Q/U$ mixing terms). Simulations include noise as it is the limiting factor of the efficiency of the correction. We have used all the detectors of the 140-GHz LFT band assumed in the previous sections and simulated one year of observations. As before we evaluate residual IP systematics in configurations with and without the 5 degree tilt of the HWP. We have run 5 simulations varying each time the noise realization. We have observed that only 5 iterations are needed to reach the convergence between $\widehat \vecm$ and $\widehat \veclambda$. This is explained by the high degree of polarization cross-linking decoupling fairly the polarization maps from the IP parameters. It takes approximately 30 minutes with 15 processors to estimate maps for one detector at the first iteration. For the other iterations the map calculation is much quicker since we start with the maps obtained with the previous iterations as initial conditions of the map-maker.

In order to isolate the effect of the $4f_{\rm HWP}$ IP terms only on the recovered Stokes $Q$ and $U$ maps after convergence of our correction method, we subtract the recovered IP terms $\widehat P_{it}$ from noiseless data, project those using the map-making operator, and subtract the input maps: 
\begin{equation}
\delta \vecm = \sum_i W_i \left[[{\cal M}^{\rm Eff}_i]^tN_i^{-1}{\cal M}^{\rm Eff}_i\right]^{-1} [{\cal M}^{\rm Eff}_i]^tN_i^{-1} (\vecd_{i}^{\rm wo/n} - \widehat \vecP_i) - \vecm
\label{eq:deltmc}
\end{equation}
with
\begin{equation}
d_{it}^{\rm wo/n} = {{\cal M}}_{itpc} \,m_{pc}^{\rm corr}.
\end{equation}
We do not include the mean intensity removal here since we focus on the polarization map residuals. The errors on the estimation of $\widehat P_{it}$ are limited by noise in input simulations. We have rescaled the input maps $m_{pc}$ to produce $m_{pc}^{\rm corr}$ in order to eliminate the effect of imperfect polarization efficiency and Q/U mixing, which are actually included in our simulations via perturbations of the Mueller matrix, in order to isolate the effect of the IP only.

\begin{figure}
  \centering
  \includegraphics[width=0.7\columnwidth]{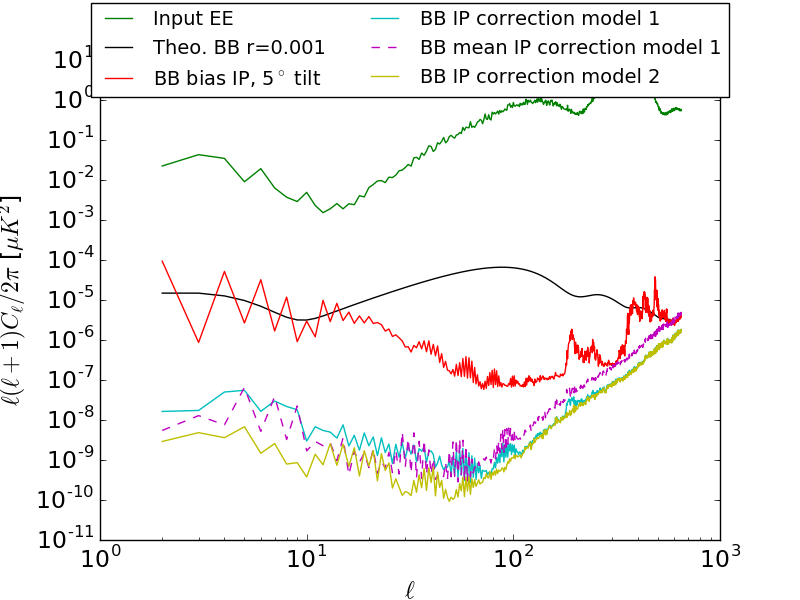}
  \caption{Residual polarization $B$-mode power spectrum calculated from the leakage maps $\delta m$ (defined in Equation~\ref{eq:deltmc}) induced by the IP HWP imperfection parameters after applying correction with a modified map-making. Power spectra amplitudes after correction are extrapolated to 3-year observation (from 1-year simulations, see text for details). Simulations combine all the detectors observing in the LFT 140-GHz band as envisaged for \LB and we assume a 5$^\circ$ tilt of the HWP rotation axis with respect to the center of the focal plane. The cyan curve is the power spectrum the residual spectrum after correction estimating all the parameters $\lambda^{(q)}_i$ individually (model 1, $\delta C^{BB}_{\ell;1}$ in the text), the yellow curve the power spectrum after correction estimating with the more constrained model where the parameters $\lambda^{(q)}$ are estimated (model 2). Spectra are the average of spectra calculated from 5 realisations. The dashed curve is the power spectrum of the mean residual map from the 5 realisations after rescaling to 3-year observations $\delta C^{BB}_{\ell;2}$ (see text for details). The red curve is the power spectrum of residual IP without correction (nearly identical for 1-year or 3-year observations). The rise of power in spectra after correction at high $\ell$ is due to the $2f_{\rm HWP}$ imperfections which are not mitigated by the method.}
    \label{Fig:SpCorr}
\end{figure}

The contribution to the $BB$ power spectrum of the $4f_{\rm HWP}$ residual effect is estimated from the residual maps $\delta m$ after convergence of the parameters. They are shown in Figure~\ref{Fig:SpCorr} for the more pessimistic case of 5$^\circ$ shift, after averaging the power spectra from the 5 realisations.
We calculated the residual power spectra for the two approaches, with the individual detector determination of $\lambda$, or the more constrained model linking the parameters. We observe a very important reduction of the residual IP effect. We have extrapolated the level of residuals for a three year observation time by rescaling the residual power spectra obtained for a one year observation time, dividing the power spectra by a factor of 3. We expect that this operation can be done because the IP parameter estimation uncertainties are noise dominated, and not systematics dominated. The residuals from IP in the maps after correction should then scale with the map noise RMS, and the power spectrum of residuals with the map noise variance. We have verified with our 5 simulations that the mean of the amplitude of the residual at the map level is compatible with 0, as well as the errors on the $\lambda$ parameters. In that case the power spectrum scales with the noise variance, and hence with the number of surveys. We have verified this by comparing the averaged $B$-mode power spectrum (over the 4 realizations numbered by $k$) with the power spectrum of the averaged map: $\delta C^{BB}_{\ell;1}=\frac{1}{4}\sum_{k=1}^4\frac{1}{3}\langle|\delta a^{B}_{\ell m;k}|^2\rangle$ and $\delta C^{BB}_{\ell;2}= \frac{1}{12}\langle| \sum_{k=1}^4 \delta a^{B}_{\ell m;k}|^2 \rangle$, respectively, where $\langle |x_{\ell m}|^2\rangle \equiv \frac{1}{2\ell+1}\sum_m |x_{\ell m}|^2$. We would observe $\delta C_{\ell;2} > \delta C_{\ell;1}$ in case of dominant systematic contributions to the residuals for different realizations, which is not the case as can be seen in the Figure. We have observed a very rapid convergence with negligible changes of the residual IP power spectra after 5 iterations. The bias $\delta r$ is estimated from the residual averaged IP power spectra $\delta C^{BB}_{\ell;1}$ after correction and results are presented in Table~\ref{tab:drIP} for the two correction methods. The correction is very effective with a reduction of power by about 3 orders of magnitude. Another order of magnitude is gained using the model linking the $\lambda_i$ parameters for different detectors (model 2). Such large improvement using model 2 instead of model 1 is expected because of the non uniform dependence of $\lambda$ on $\Theta$ reducing the uncertainties of $\lambda_i$ for detectors near the HWP optical axis.

Let us focus now on the accuracy of the estimation of the $\lambda$ parameters, and on the theoretical predictions of the propagation of those uncertainties on the level of the remaining leakage in the $Q$ and $U$ maps after correction.
The quantities 
\begin{equation}
  {\rm Var}\left\{\veclambda_i|\vecm\right\} = \sigma_n^2 [T_i^t T_i]^{-1},\label{Eq:Varlamb1}
\end{equation}
for individual detectors and
\begin{equation}
  {\rm Var}\left\{\veclambda|\vecm\right\} = \left[\sum_i Z_i^tN_i^{-1}Z_i\right]^{-1},\label{Eq:VarlambAll}
\end{equation}
for the global parameters, would give the variance of the IP parameters under the condition of fixed $m$ (and in particular $I$ in the expression of $T$ or $U$) with $\sigma_n^2$ the white noise variance per sample (we remind that the 1/f part is irrelevant since the templates are varying at $4f_{\rm HWP}$). This estimation would then neglect the coupling between the estimated IP and Stokes map parameter, and we expect some level of correlation since the templates are built from the recovered intensity $I$ map. 
The calculated uncertainties (square root of the parameter variances) are $\sigma(\lambda^{(1,2)}_i) = 6.0 \times 10^{-6}$ with very little variation from detector to detector (1 year full time observations)~\footnote{The other parameters $\lambda^{(3)}$ and $\lambda^{(4)}$ provide a fit of the monopole signal and do not carry significant additional information on the phase of the IP parameters due to the low monopole amplitude in simulations (the situation will be different will real data)}. The correlation between the four parameters is negligible as expected. Using those number we can predict roughly the residual bias $\delta r$ by simply calculating
\begin{equation}
    \delta r_{\rm corr} \leq \frac{2}{N_{\rm det}}{\rm Var}\left\{\lambda_i^{(1)}|m\right\} \frac{\delta r_{\rm ref}}{{\epsilon_{\rm ref}}^2}
\end{equation}
using any reference detector and under the following hypothesis: 1. quadratic scaling of $\delta r$ with $\epsilon_i$ or $\lambda_i$; 2. no coupling between IP and Stokes map parameters.
Using for example the estimated power spectrum from one of the detector displayed in Figure~\ref{SpSigleak}, we calculated the corresponding $\delta r_{\rm ref}$ maximizing the likelihood function on $r$ and we find $\delta r_{\rm corr}\leq 2 \times 10^{-7}$ (for 3 year observations). This value is very significantly below, by a factor $\approx 6$, our estimation from simulations in Table~\ref{tab:drIP} for model 1. We attribute this discrepancy to the correlation of errors between the map estimate (which is used for building the IP template) and the IP parameters. Large number of Monte Carlo simulations would be necessary for a detailed study of the origin of the discrepancy.

We do not observe significant bias on the estimated parameters from our limited number of realisations, however more realisations are necessary for a detailed study. We have observed that neglecting the effect of the monopole, i.e. fitting only for $\lambda^{(1)}$ and $\lambda^{(2)}$ (and not $\lambda_{(3)}$ and $\lambda^{(4)}$), has a significant impact on the final results, even in the case where we have only the Galactic monopole in simulations.

Our final comment concerns multi-wavelength analysis. Since the estimation of IP parameters is limited by noise (in absence of other systematics effect that could potentially couple with this estimation), the amplitude of the contamination in the different frequency bands should be randomly distributed around 0. Then under the hypothesis that the 140-GHz channel of LFT is one of the channels with the largest effect given the nominal \LB configuration since it is at the edge of the HWP bandwidth, the residual bias combining different frequency channels should average down and be reduced. This study is postponed to future publications.

\begin{table}[ht!]
 \centering
			\begin{tabular}{lp{0.88\textwidth}}
                \hline
                $i$ & detector number index.\\
                $t$ & time index.\\
                $p$ & single map pixel index. \\
                $q$ & template component index. \\
                $c$ & Stokes parameter component: $c=0 \rightarrow I$; $c=1,2 \rightarrow Q,U$.\\
                $\widehat .$ & estimated quantity.\\
				$\epsilon_i$ & Mueller matrix element amplitude varying at 4$f_{\rm spin}$ $M^{(4f)}_{QI}(\Theta_i,\Phi_i)$ \par ($ = M^{(4f)}_{UI}(\Theta_i,\Phi_i))$. \\
                $\epsilon_{\rm ref}$ & same as previous entry but for a reference detector. \\
				$\lambda^{(q)}_i$ & parameter multiplying the template $T_{itq}$. It is related to the IP parameters by $\epsilon_i=\sqrt{(\lambda^{(1)}_i)^2+(\lambda^{(2)}_i)^2}$.  \\
    			$\lambda^{(q)}$ & parameter multiplying the template $Z_{itq}$. It verifies $\lambda^{(q)}=\lambda^{(q)}_{i_{\rm ref}}$, with $i_{\rm ref}$ the index of the reference detector.\\
                $T_{itq}$ & timestream template induced by HWP imperfections built from the intensity map.\\
                $Z_{itq}$ & timestream template induced by HWP imperfections built from the intensity map.\\
                $m_{pc}$ & element of map vector related to the Stokes component $c$ for pixel $p$.\\
                $d_{it}$ & element of data timestream.\\
                $n_{it}$ & element of noise timestream.\\
                ${\cal M}_{itpc}$ & matrix element expressing the contribution of the Stokes parameter map $c$ at pixel $p$ to the observation data at time $t$ for detector $i$. It incorporates the HWP Mueller matrix $M$.\\
                ${\bar{\cal M}}_{itpc}$ & same as ${\cal M}_{itpc}$ but with the IP terms set to 0.\\
                ${\cal M}^{\rm Eff}_{itpc}$ & same as ${\cal M}_{itpc}$ but incorporating an ideal HWP Mueller matrix $M^{\rm Ideal}$.\\
                $N_i$ & noise covariance matrix of dimension number of sample square.\\
                $P_{it}$ & $=T_{itq}\lambda_i^{(q)}$ or $=Z_{itq}\lambda^{(q)}$ dependng on the model used.\\
                $W_i$ & weights multiplying maps.\\
                ${\overline{\widehat I_i}}$ & mean of the estimated intensity map for detector $i$.\\
                $\vecu$ & single map size vector containing 1 as components for each pixel. \\
                \hline
                & We use boldface for 1D vectors: $m_{pc}\rightarrow {\bf m}$; $d_{it}\rightarrow {\bf d}_i$; $n_{it}\rightarrow {\bf n}_i$; $\lambda^{(q)}_i\rightarrow {\bf \veclambda}_i$; $\lambda_i\rightarrow {\bf \veclambda}$; $P_{it}\rightarrow \vecP_i$, \par and compact notation for other higher dimension matrices: ${\cal M}_{itq} \rightarrow {\cal M}_i$, Mueller/pointing matrix for detector $i$; $T_{itq}\rightarrow T_i$; $Z_{itq}\rightarrow Z_i$.\\
				\hline
			\end{tabular}
\caption{Definition of indices, variables and notations in Section~\ref{sec:correction}.
\label{tab:variables}
}
\end{table}

\section{Conclusions}

We have estimated and modeled the effect of leakage from unpolarized intensity to polarization induced by HWP imperfections
for a \LBm-type satellite mission with a HWP used as first optical element, and for the \LB observation strategy for the measurement of CMB $B$-modes. We propagate to CMB polarization maps the effect of imperfections using a Mueller matrix formalism which is meant to be general, however we use as test-case semi-analytical predictions from EM propagation calculators for a multi-layer HWP including anti-reflection coating (assuming ideal fabrication). The predicted Mueller matrix coefficients used for this work are decomposed in 0th, 2nd and 4th harmonics of the rotation frequency. The amplitudes of the 4th harmonics of the instrumental polarization parameters $M_{QI}$ and $M_{UI}$ is of the order of $10^{-5}$ to $10^{-4}$ depending on the line of sight incident angle on the HWP.

We have shown analytically that the amplitude of the IP leakage in the recovered polarization maps for one individual detector is closely related to the spin-2 cross-linking parameters $<\cos 2\psi>_p$ and $<\sin2\psi>_p$ in each pixel $p$ for the corresponding detector location and the relative phase between the sine and cosine terms depend on the azimuthal angle of the detector location around the HWP axis. Consequently, some level of cancellation of the effect happens when combining different detectors at different locations in the focal plane. In particular we show that the cancellation is maximal between two detectors with azimuthal angle difference of 90$^\circ$.
We have run simulations for the whole LFT 140-GHz band of \LB with or without a tilt of the HWP rotation axis. The cosmological dipole is the main source of leakage and CMB anisotropies and foregrounds have small contributions. We have estimated $B$-mode power spectrum of the leakage and an induced bias on the tensor-to-scalar ratio $\delta r$ of $1.36 \times 10^{-4}$ without HWP tilt and $1.47 \times 10^{-3}$ for a 5$^{\circ}$ tilt, for an averaged IP parameter across detectors $\epsilon$, the amplitude of $M_{QI}^{(4f)}$ (which we assume equal to $M_{UI}^{(4f)}$), of 2.23 $\times 10^{-5}$ and 4.38 $\times 10^{-5}$, respectively. In case a different HWP model is used, this result can be rescaled detector by detector for any amplitude of $\epsilon$ since the amplitude of the leakage is proportional to $\epsilon$ (the bias on $r$ is then quadratic with respect to $\epsilon$). The CMB monopole (not included in simulation) is expected to have a large impact but appears as a pure 4f component, i.e single line for a constant rotating HWP, which can be subtracted without harm on the CMB signal by performing a joint fitting. However, interplay with other possible systematic effects such as detector nonlinearities, gain fluctuations or loading fluctuations would induce a broad band contribution and could potentially lead to spurious $B$-mode signal. We verify that the 2f component has small impact on the polarization maps.

We have provided and implemented a maximum likelihood method to estimate the HWP parameters from the flight data and to correct the leakage from the polarization maps. This is a generalized map-making method allowing the joint estimation of the Stokes parameter maps along with the two amplitudes of the Mueller matrix coefficients at 4f (we have two amplitude parameters since the amplitude and phase are unknown), $\lambda^{(1)}$ and $\lambda^{(2)}$ for cosine and sine components, as well as the two amplitudes of a constant 4f signal, since the monopole signal is a free parameter as it is only approximately measured by the experiment. The solutions for the two sets, i.e. the maps and the IP parameters, are obtained iteratively. The final maps are naturally corrected from the leakage at convergence. We have implemented two different HWP models in the correction method, one with free $\epsilon_i$ (or $\lambda^{(1,2)}_i$) parameters for each detector $i$, another using an empirical scaling law of $\epsilon$ depending on the incident angle on the HWP. In this last case we fit for only four global IP parameters.

We have evaluated the performance of the approach with the help of simulations of observations of the \LB detector in the LFT 140-GHz band including the sky signal and nominal noise. We have shown that the amplitude of the sine and cosine components of the 4f Mueller matrix coefficients can be estimated with a noise statistics-limited precision of $6\times 10^{-6}$ per detector for the most relaxed model. The residual effect after correction on the final polarization maps for the full LFT 140-GHz band leads to the tensor-to-scalar ratio $\delta r=1.84\times 10^{-7}$, which is negligible given the target of \LB and the allocated budget for systematics. The correction leads to $\delta r=1.2\times 10^{-6}$ using the constrained model.

After correction, the remaining leakage amplitude due to uncertainties in the parameter estimation is a zero mean quantity. Consequently we expect that the remaining effects in different frequency band maps are uncorrelated. Then after combining those maps with component separation methods, we expect that the total residual leakage on the final CMB polarization map will be of lower amplitude than in each individual channel leading to an even smaller bias on $r$. Multi-frequency simulations and a full pipeline of analysis is postponed to a future study.

\appendix
\section{From Jones to Mueller Matrix Perturbations}
\label{Ap:A}

The objective of this note to relate the perturbations in the Jones matrix formalism to the ones in the Mueller matrices that are used in this paper.

First, we define the Stokes vector from the electric fields, i.e. the Jones vector elements, as 
\begin{align}
I&=\langle |E^{(x)}|^2 + |E^{(y)}|^2\rangle\\
Q&=\langle |E^{(x)}|^2 -|E^{(y)}|^2 \rangle\\
U&=2\langle {\rm Re}\{E^{(x)}{E^{(y)}}^*\}\rangle\\
V&=2\langle {\rm Im}\{E^{(x)}{E^{(y)}}^*\}\rangle.
\end{align}
Second, we define $M^{\rm HWP}(\alpha)$ and $J^{\rm HWP}(\alpha)$ as the Mueller and Jones matrices of the HWP, respectively, in a frame defined by the HWP ordinary and extraordinary axis for a given incident angle $\Theta$ (not specified in the relations in this appendix were we assume this angle is fixed). We define $\alpha$ as the orientation angle of the HWP with respect to the observation frame (the focal plane), and can be assimilated to $\alpha=\rho-\psi$. The angle $\alpha$ can be viewed as the azimuthal observation angle around the HWP axis, so the dependency of the Jones or Mueller matrix coefficients with respect to $\alpha$ reflects the change of the polarization state induced by HWP imperfections depending on the line of sight perspective even in the fixed frame (with respect to the HWP). The Stokes vector $\vec{S}$ and the electric field transform using the Mueller and Jones matrices, respectively, as. 
\begin{eqnarray}
\vec{S}_{out} &=& R(2\psi_0)\,R^{-1}(2\alpha)\,M^{\rm HWP}(2\alpha)\,R(2\alpha) \,R(2\psi)\, \vec{S}_{in}, \\
E_{out} &=& R(\psi_0)\,R^{-1}(\alpha)\,J^{\rm HWP}(\alpha)\,R(\alpha)\,R(\psi)\,E_{\rm in}.
\end{eqnarray}
To be explicit, the HWP Mueller matrix $M$ used throughout the paper (Eq.~\ref{eq:modelobs}) is equivalent to 
\begin{equation}
M(\alpha) = R(-2\alpha) M^{\rm HWP}(\alpha) R(2\alpha),
\label{eq:relMuellerref}
\end{equation}
and is the HWP Mueller matrix in the focal plane frame.
The Mueller matrix elements of interest are related to those of the Jones matrices via 
\begin{align}
M^{\rm HWP}_{QI} &= \frac{1}{2} \Big(|J_{11}^{\rm HWP}|^2-|J_{22}^{\rm HWP}|^2+|J_{12}^{\rm HWP}|^2-|J_{21}^{\rm HWP}|^2\Big), \\
M^{\rm HWP}_{UI} &= {\rm Re} \Big\{J_{11}^{\rm HWP}{J_{21}^{\rm HWP}}^* + J_{12}^{\rm HWP}{J_{22}^{\rm HWP}}^*\Big\}.
\end{align}

Let us now apply this relation to the specific form of $J^{\rm HWP}(\alpha)$, representing the small perturbation from an ideal HWP, as 
\begin{equation}
  J^{\rm HWP}(\alpha) = 
\begin{pmatrix}
1 + \delta_{11}(\alpha) &  \delta_{12}(\alpha)\\  \delta_{21}(\alpha) & -1 + \delta_{22}(\alpha)
\end{pmatrix}
\end{equation}
The corresponding Mueller matrix elements are
\begin{align}
M^{\rm HWP}_{QI} &= {\rm Re}\Big\{\delta_{11} + \delta_{22}\Big\} + \frac{1}{2} \Big(|\delta_{11}|^2 + |\delta_{12}|^2 - |\delta_{21}|^2 - |\delta_{22}|^2\Big) \\
&\approx {\rm Re}\Big\{\delta_{11} + \delta_{22}\Big\}\label{eq:m1}, \\
M^{\rm HWP}_{UI} &= {\rm Re}\Big\{{\delta_{21}}^* - \delta_{12} + \delta_{11}{\delta_{21}}^* + \delta_{12}{\delta_{22}}^*\Big\} \\ 
& \approx {\rm Re}\Big\{\delta_{21} - \delta_{12}\Big\}\label{eq:m2}.
\end{align}
The last approximation is to neglect the quadratic terms. 
The perturbation coefficients can be expanded into harmonics of the rotation frequency:
\begin{equation}
	\delta_{pq} = \sum_n \gamma_{pq}^{(n)}\exp(in\alpha).
\end{equation}
Consequently, due to the (quasi) linear relation between Mueller and Jones matrix coefficient, Mueller matrix coefficients can be expanded in the same way:
\begin{align}
M^{\rm HWP}_{QI} &= \sum_n A_{1;n}\cos(n\alpha + n\phi_{1;n})\\
M^{\rm HWP}_{UI} &= \sum_n A_{2;n}\cos(n\alpha + n\phi_{2;n}),
\end{align}
where the (real) amplitudes $A_{i;n}$ and the phases $\phi_{i;n}$ can be easily related to $\gamma_{pq}^{(n)}$,
without coupling between modes in the Mueller matrix, i.e. a single mode of frequency $f_n$ in Jones matrix will give a single mode of the same frequency in the Mueller matrix.

We now calculate the IP coefficients of the Mueller matrix in the focal plane coordinates $M$ as used in the paper. Using Equations~\ref{eq:m1} and~\ref{eq:m2} combined with~\ref{eq:relMuellerref} we obtain:
\begin{align}
M_{QI} &= \frac{1}{2}\sum_n \Big[ A_{1;n}\Big(\cos\big([n-2]\alpha+n\phi_{1;n}\big) + \cos\big([n+2]\alpha+n\phi_{1;n}\big)\Big) \\ 
& + A_{2;n}\Big(\sin\big([n-2]\alpha+n\phi_{2;n}\big) - \sin\big([n+2]\alpha+n\phi_{2;n}\big)\Big) \Big] \\
M_{UI} &= \frac{1}{2} \sum_n \Big[ A_{1;n} \Big(\sin\big([n+2]\alpha+n\phi_{1;n}\big) - \sin\big([n-2]\alpha+n\phi_{1;n}\big)\Big) \\ 
& + A_{2;n} \Big(\cos\big([n-2]\alpha+n\phi_{2;n}\big) + \cos\big([n+2]\alpha+n\phi_{2;n}\big)\Big) \Big].
\end{align}
From the previous relations, we see that the $4f_{\rm HWP}$ IP leakage can originate from $2f_{\rm HWP}$ ($n=2$) and $6f_{\rm HWP}$ ($n=6$) perturbation modes in the Jones matrix. We can show that the $n=2$ modes are always leading to a phase difference of $\pi/2$ between the two IP terms $M_{QI}^{(4f)}$ and $M_{UI}^{(4f)}$, as assumed in this paper and as verified by RCWA simulations, while the $n=6$ terms induce the opposite phase difference. A mix of the two modes could produce any phase difference. Simulations indicate that the $n=2$ modes greatly dominate over the $n=6$ modes, so the phase difference relation $\phi^{(4f)}_{QI} = \phi^{(4f)}_{UI} + \frac{\pi}{2}$ holds with a very high precision although it is not a strict relation.

\acknowledgments

The authors thank Luca Pagano and Francesco Piacentini for useful comments on the draft which helped improving the quality of the paper.
GP acknowledge the International Research Fellow of Japan Society for the Promotion of Science (Invitational Fellowships for Research in Japan (Short-term)).
This work has also received funding by the European Union’s Horizon 2020 research and innovation program under grant agreement No 101007633 CMB-Inflate. GP and HI acknowledge the RECTOR program in Okayama University.
This work is supported by JSPS KAKENHI Grant Numbers JP18KK0083 and JP23H00107, and by the JSPS Core-to-Core Program.
\textit{LiteBIRD} (phase A) activities are supported by the following funding sources: ISAS/JAXA, MEXT, JSPS, KEK (Japan); CSA (Canada); CNES, CNRS, CEA (France);
DFG (Germany); ASI, INFN, INAF (Italy); RCN (Norway); AEI (Spain); SNSA, SRC (Sweden); and NASA, DOE (USA).
Some of the results in this paper have been derived using the HEALPix package.

\bibliographystyle{JHEP}
\bibliography{BiblioInstruPolar}





\end{document}